\documentclass[11pt,a4paper]{article}
\usepackage{jheppub}
\usepackage{amsfonts}
\usepackage{amssymb}

\usepackage{graphicx}

\newcommand{\bea}{\begin{eqnarray}}
\newcommand{\eea}{\end{eqnarray}}
\newcommand{\bean}{\begin{eqnarray*}}
\newcommand{\eean}{\end{eqnarray*}}
\newcommand{\nn}{\nonumber\\}
\newcommand{\Sl}{\sum\limits}

\allowdisplaybreaks
\title{On Primary Relations at Tree-level in String Theory and Field Theory}

\author[]{Qian Ma}
\author[]{Yi-Jian Du\footnote{Corresponding author}}
\author[]{Yi-Xin Chen}
\affiliation[]{Zhejiang Institute of Modern Physics, Zhejiang University,\\
Hangzhou, 310027, P. R. China}
\emailAdd{mathons@zju.edu.cn}
\emailAdd{yjdu@zju.edu.cn}
\emailAdd{yxchen@zimp.zju.edu.cn}

\abstract{By the use of cyclic symmetry, KK relations and BCJ relations, one can reduce the number
of independent $N$-point color-ordered tree amplitudes in gauge theory and string theory from $N!$
to $(N-3)!$. In this paper, we investigate these relations at tree-level in both string theory and
field theory. We will show that there are two primary relations. All other relations can be
generated by the primary relations. In string theory, the primary relations can be chosen as
cyclic symmetry as well as either the fundamental KK relation or the fundamental BCJ relation.
In field theory, the primary relations can only be chosen as cyclic symmetry and the fundamental BCJ
relation. We will further show a kind of more general relation which can also be generated by the primary
relations. The general formula of the explicit minimal-basis expansions for  color-ordered open string
tree amplitudes will be given and proven in this paper.
}
\keywords{Gauge symmetry, QCD}
\arxivnumber{arXiv:1109.0685}
\begin{document}
\maketitle
\section{Introduction}

The relations among scattering amplitudes play an important role in understanding gauge field theory
and quantum gravity. In gauge field theory, color-ordered amplitudes at tree-level have been shown
to satisfy many relations. These relations provide constraints on amplitudes. With these relations,
one can expand any $N$-point color-ordered tree amplitude by a minimal-basis of $(N-3)!$ amplitudes.
The first relation is the cyclic symmetry with which one can reduce the number of independent amplitudes from
$N!$ to $(N-1)!$. The second one is the Kleiss-Kuijf relation\cite{Kleiss:1988ne}(KK relation). With
the KK relations, one can express the $(N-1)!$ amplitudes by only $(N-2)!$ independent amplitudes. The
coefficients in front of the amplitudes in cyclic symmetry and KK relations are independent of kinematical factors
$s_{ij}=(k_i+k_j)^2$. The third one is the Bern-Carrasco-Johansson relation\cite{Bern:2008qj}(BCJ relation),
with which the number of independent amplitudes can be further reduced down to $(N-3)!$.
Being different from the cyclic symmetry and the KK  relations, BCJ relations are highly nontrivial
relations, i.e., the coefficients of the amplitudes in BCJ relations are functions of kinematic factors $s_{ij}$.
In quantum gravity, there are Kawai-Lewellen-Tye(KLT) relations\cite{Kawai:1985xq}
that express graviton amplitudes in terms of products of two gluon amplitudes. As in the BCJ relations,
there are also nontrivial coefficients in the  KLT relations.

All these amplitude relations have been studied in both field theory and string theory. In string
theory, the cyclic symmetry is resulted by worldsheet conformal symmetry. Both KK and BCJ relations
come from the so-called monodromy relations which can be derived by deforming the contour of
worldsheet integrals\cite{BjerrumBohr:2009rd, Stieberger:2009hq, BjerrumBohr:2010zs}. KK relations
are the real part relations, while BCJ relations are the imaginary part relations. KLT relations are
also monodromy relations\cite{Kawai:1985xq,BjerrumBohr:2010hn}. Though the coefficients of the
amplitudes in KK relations in field theory are trivial, we should notice that these coefficients
in KK relations in string theory are nontrivial functions(cosine functions) of kinematic factors.
Thus, in string theory, the only relation with trivial coefficients is the cyclic symmetry.

After taking the field theory limits from string theory, we get the amplitude relations in field
theory. However, for the consideration of consistency, pure field theory proofs are also necessary.
In field theory, the KK relations were first proven via new color decomposition\cite{DelDuca:1999rs},
while the fundamental BCJ relations were proven in \cite{Feng:2010my, Jia:2010nz} by using BCFW
recursion\cite{Britto:2004ap, Britto:2005fq}\footnote{ In \cite{Tye:2010kg}, BCJ relations were
 considered  by using Schouten identity.}. In \cite{Feng:2010my, Jia:2010nz}, it was stated
 that one can use a set of fundamental BCJ relations to solve the minimal-basis expansion out.
 But it seems impossible to prove the general formula of explicit minimal-basis
 expansion\cite{Bern:2008qj} in this way. This is because the coefficients before the amplitudes
 are too complicated. The general formula of explicit minimal-basis expansion in field theory was
 proven in \cite{Chen:2011jx} via a set of relations called general BCJ relations which are field
theory limits of the BCJ relations in string theory(See \cite{BjerrumBohr:2009rd}). KLT relations
 in field theory have been proven in \cite{BjerrumBohr:2010ta, BjerrumBohr:2010zb,
 BjerrumBohr:2010yc, Feng:2010br} by BCFW recursion. BCJ relations play an important role in the
 field theory proof of the KLT relations.

Another representation of BCJ relation is referred as Jacobi-like identity among the kinematic numerators
\cite{Bern:2008qj}. Works in this representation can be found in, e.g., \cite{Bern:2010ue,
 Bern:2010yg}. Recent researches on BCJ numerator can be found in \cite{Mafra:2011kj, Monteiro:2011pc}.
The relations in heterotic string theory was studied in \cite{Tye:2010dd}. Many works on amplitude
relations via pure spinor string can be found in\cite{Mafra:2009bz, Mafra:2010gj}, \cite{Mafra:2011kj},
\cite{Mafra:2011nv, Mafra:2011nw}. It is interesting that the KK and BCJ relations are not only hold
in gauge field theory but also in other cases, e.g., the KK relations also hold for amplitudes with
gluons coupled to gravitons\cite{Chen:2010ct}, while the BCJ relations were also suggested to
be hold for gluons coupled to matters\cite{Sondergaard:2009za}. In \cite{Du:2011js},  the KK and BCJ relations
in color dressed scalar theory were proven by BCFW recursion with nontrivial boundary
\cite{Feng:2009ei, Feng:2010ku}. KLT relations are also extensive
relations and can be used in many cases\cite{Bern:1999bx}. The KLT relations for pure gauge amplitudes
are proven in \cite{Du:2011js}. The extension of the amplitude relations
to loop-level can be found in \cite{Bern:2010ue, Bern:2010yg, BjerrumBohr:2011xe, Feng:2011fja}.
A dual formula of color decomposition was proposed in \cite{Bern:2011ia}.

Although there have been a lot of studies on amplitude relations, there is an important thing
that should be emphasized: The general BCJ relations which were used to prove the minimal-basis
expansion in \cite{Du:2011js} are redundant ones if we consider the fundamental BCJ relations as
the primary relations. This is because the minimal-basis expansion is the solution of the general
BCJ relations\cite{Chen:2011jx} and it can
also be solved out from a set of  fundamental BCJ relations\cite{Feng:2010my,
 Jia:2010nz}.  It is not apparent
to extend this statement to the KK relations in field theory. Can we also generate KK relations
by some primary relations? If we consider all the cyclic symmetry, KK relations and BCJ
relations together, we have a further question: Can we generate all these relations by some primary
relations? In fact, as we have mentioned above, in string theory, both KK and BCJ
relations come from monodromy. Since they are just the real part and imaginary part of
the same monodromy relation, they have fairly equal status in string theory. Thus we speculate
that the KK relations can also be generated by some primary relations. In this paper,
we will treat all the cyclic symmetry, KK and BCJ relations together.
We will show that all these relations in string theory can be generated by two primary relations.
One primary relation is the cyclic symmetry while the other one
can be chosen as either the fundamental KK relation or the fundamental BCJ relation. This argument
can be extended to field theory. In field theory, all the KK and BCJ relations can be generated
by the fundamental BCJ relation and the cyclic symmetry. The difference from the string theory case
is that the primary relations cannot be chosen as the $U(1)$-decoupling identity(field theory limit of
fundamental KK relation in string theory) in field theory. The discussions in string theory
can be achieved by the following steps(See \eqref{steps}):
\bea\label{steps}
&&\begin{array}{c}
 \begin{array}{c}
    cyclic \\
     fundamental~KK(BCJ)
   \end{array}
   \Bigr\}\Rightarrow~fundamental~BCJ(KK)
   \\
\end{array}
\Biggr\}\Rightarrow~U(1)-like~decoupling\nn
&&\Rightarrow generalized~U(1)-like~decoupling\Rightarrow~KK-BCJ\Rightarrow
\Biggl\{\begin{array}{c}
  KK \\
  BCJ
\end{array}
\eea
Firstly, we will write the KK and the BCJ relations in string theory into the  monodromy
relations named KK-BCJ relations. The simplest KK-BCJ relation is the $U(1)$-like decoupling
identity(the real part gives the fundamental KK relation, while the imaginary part gives the
fundamental BCJ relation). Secondly, we will show that the fundamental KK(BCJ) relation can
be generated by the cyclic symmetry and the fundamental BCJ(KK) relation. Thus we can choose
 the cyclic symmetry as well as either one of the fundamental KK relation and the fundamental
 BCJ relation as the primary relations to generate the $U(1)$-like decoupling identity. Thirdly, we
will generate a kind of relation named
generalized $U(1)$-like decoupling
identity\footnote{The
 generalized $U(1)$-decoupling identity in field theory was given in \cite{Du:2011js}. This
 relation is just the field theory limit of the real part of the generalized $U(1)$-like decoupling
 identity in this paper.} by using the $U(1)$-like decoupling identity. At last, we will
 show the equivalence
 between generalized $U(1)$-like decoupling identities and KK-BCJ relations. Thus all the KK and BCJ
 relations can be generated by only two primary relations. The field theory results can be
 obtained by taking field theory limits carefully.

In our discussions of this paper, various linear combinations of amplitudes accompanied by
momentum kernels are useful. We will use the linear combinations of amplitudes to give a kind of more
general relation which is mentioned as general monodromy relation in this paper. The generalized $U(1)$-like
decoupling identities and the KK-BCJ relations can be considered as special cases of this kind of
relation. The general monodromy relations can also be generated by the primary relations. After
 taking the field theory limits, we get the corresponding relations in field theory.

As we have mentioned above, the minimal-basis expansion at tree-level for color-ordered amplitudes
can be solved from the fundamental BCJ relations
 \cite{Feng:2010my, Jia:2010nz}, but it seems impossible to give the general formula in this way.
 Though the general formula of minimal-basis expansion for color-ordered pure-gluon tree amplitudes
 in field theory has been conjectured in \cite{Bern:2008qj} and proven in \cite{Chen:2011jx}, the
  corresponding expression has not been found in string theory.
In this paper, we will derive the explicit minimal-basis expansion for color-ordered open string
tree amplitudes.

The structure of this paper is as follows. In section \ref{section_2}, we will give a new defined
momentum kernel which will be useful in this paper. We will rewrite the monodromy relations by
momentum kernels and show an $(N-2)!$-formula of the color decomposition for open string tree
amplitudes via KK-BCJ relation.   In section \ref{section_3}, we will show all the KK and BCJ
relations can be generated by the primary relations. In section \ref{section_4}, we will extend
the combination of amplitudes to give the general monodromy relation. In section \ref{section_5},
we will derive the minimal-basis expansion for color-ordered open string tree amplitudes. A summary
of our conclusion will be given in section \ref{section_6}.

\section{A new defined momentum kernel and the $(N-2)!$-formula of color decomposition in string theory }\label{section_2}

Momentum kernel has been suggested in \cite{BjerrumBohr:2010hn} and it plays an important role
 in understanding monodromy relations.
In this section, we will give discussions on a new defined momentum kernel which will be
useful in this paper. In subsection \ref{sub_momentum_kernel}, we will provide the definition
and useful properties of this momentum kernel. By the use of this  momentum
kernel we will rewrite the monodromy relations in string theory.

Using the new defined momentum kernel in string theory, one can express the KK and BCJ relations
 in a complex formula(KK-BCJ relations). With this formula, any amplitude can be expressed
 as a linear combination
 of $(N-2)!$ KK basis accompanied by appropriate complex momentum kernels. Thus we expect that there
is also an $(N-2)!$-formula of color decomposition in string theory as in the case of field
theory\cite{DelDuca:1999rs}. In subsection \ref{sub_color_decomposition}, we will show the
$(N-2)!$-formula of color decomposition in string theory by a new defined commutator.

\subsection{A new defined momentum kernel in string theory}
\label{sub_momentum_kernel}

In this paper, the momentum kernel is defined as
\bea\label{kernel}
\mathcal{P}_{\{\sigma\},\{\tau\}}
=\exp\left[-2i\pi\alpha'\Sl_{i,j}k_i\cdot k_j\theta(\sigma^{-1}(i)-\sigma^{-1}(j))\theta(\tau^{-1}(j)-\tau^{-1}(i))\right],
\eea
where
\bea
\theta(x)=\Bigl\{
            \begin{array}{cc}
              1 & (x>0) \\
              0 & (x\leq 0) \\
            \end{array}
\eea
Here, $\sigma$ and $\tau$ are two permutations of the external legs.
 We denote the external leg at the $j$-th location in a given permutation $\sigma$ as $\sigma_j$.
 If $\sigma_j=i$ for some external leg $i$, we can denote $j$ as $\sigma^{-1}(i)$. For example, in
 the permutation $\sigma=3,1,2$, we have $\sigma^{-1}(3)=1$, $\sigma^{-1}(1)=2$ and $\sigma^{-1}(2)=3$.
The momentum kernel is defined as a phase factor which depends on the relative orderings of the legs in
two sets $\{\sigma\}$ and $\{\tau\}$. We take some momentum kernels with three legs as examples
\bea
\mathcal{P}_{\{1,2,3\},\{3,1,2\}}=\exp\left[-2i\pi\alpha'(k_1\cdot k_3+k_2\cdot k_3)\right],
\mathcal{P}_{\{2,1,3\},\{1,2,3\}}=\exp\left(-2i\pi\alpha'k_1\cdot k_2\right),
\mathcal{P}_{\{3,1,2\},\{3,1,2\}}=1.\nonumber
\eea
From the definition and the above examples, we can see  if the relative orderings of two legs $i$ and $j$
are different in the two permutations $\sigma$ and $\tau$, this momentum kernel gets a factor
$e^{-2i\pi\alpha'k_i\cdot k_j}$. Else, if the relative orderings of two legs in $\sigma$ and $\tau$ are same,
the momentum kernel only gets a trivial factor $1$.
For any two legs, we have a factor  $e^{-2i\pi\alpha'k_i\cdot k_j}$ or $1$. After considering all the
relative orderings in the two permutations, we get the whole momentum kernel.
The following properties of the momentum kernel will be useful in this paper:

{(i)} If the two permutations are identical, we have
\bea\label{kernel_1}
\mathcal{P}_{\{\sigma_1,\sigma_2,...,\sigma_N\},\{\sigma_1,\sigma_2,...,\sigma_N\}}=1.
\eea
{(ii)} The momentum kernel is symmetric under $\sigma\leftrightarrow\tau$
\bea\label{kernel_2}
\mathcal{P}_{\{\sigma\},\{\tau\}}=\mathcal{P}_{\{\tau\},\{\sigma\}}.
\eea
{(iii)} For a given permutation  $\tau$ which satisfies
$\tau \in P(O\{\sigma_i,\sigma_{i+1},...,\sigma_{i+j}\}
\bigcup \{\sigma_1,...,\sigma_{i-1},\sigma_{i+j+1},...,\sigma_N\})$,
i.e., $\tau$ is a permutation preserving the relative ordering of
$\sigma_i$,$\sigma_{i+1}$,...,$\sigma_{i+j}$, we have
\bea\label{kernel_3}
&&\mathcal{P}_{\left\{\sigma_1,...,\sigma_{i-1},\sigma_{i+j},...,\sigma_{i},\sigma_{i+j+1},...,\sigma_N\right\},
\left\{\tau\right\}}\nn
&=&e^{-2i\pi\alpha'\Sl_{i\leq m<n\leq i+j}k_{\sigma_i}\cdot k_{\sigma_j}}
\mathcal{P}_{\left\{\sigma_1,...,\sigma_{i-1},\sigma_i,...,\sigma_{i+j},\sigma_{i+j+1},...,\sigma_N\right\},\left\{\tau\right\}}.
\eea
{(iv)} For a given $\tau\in P(\{\gamma_1,...,\gamma_t\}\bigcup O\{\beta_1,...,\beta_s\})$, we have
\bea\label{kernel_4}
&&\mathcal{P}_{\{\gamma_1,...,\gamma_t,\alpha_1,\beta_1,...,\beta_s,\alpha_2,...,\alpha_r\},\{\tau,\alpha_1,...,\alpha_r\}}\nn
&=&e^{(-2i\pi\alpha'\Sl_{i=1}^s k_{\alpha_1}\cdot k_{\beta_i})}\mathcal{P}_{\{\gamma_1,...,\gamma_t,\beta_1,...,\beta_s,\alpha_1,...,\alpha_r\},\{\tau,\alpha_1,...,\alpha_r\}}.
\eea
{(v)} The factorization relation: if $\sigma\in P(O\{\gamma\}\bigcup O\{\beta\} \bigcup O\{\phi\})$, we have
\bea\label{kernel_5}
\mathcal{P}_{\{\gamma,\beta,\alpha,\phi\},\{\sigma,\phi\}}
=\mathcal{P}_{\{\gamma,\beta,\alpha,\phi\},\{\sigma/\{\alpha\},\alpha,\phi\}}\mathcal{P}_{\{\sigma/\{\alpha\},\alpha,\phi\},\{\sigma,\alpha,\phi\}}
=\mathcal{P}_{\{\gamma,\beta,\alpha,\phi\},\{\gamma,\sigma/\{\gamma\},\phi\}}\mathcal{P}_{\{\gamma,\sigma/\{\gamma\},\phi\},\{\sigma,\alpha,\phi\}},
\eea
where  $\gamma$, $\beta$, $\alpha$, $\phi$ denote the permutations $\gamma_1,...,\gamma_t$, $\beta_1,...,\beta_s$,
 $\alpha_1,...,\alpha_r$, $\phi_1,...,\phi_q$. $\sigma/\{\gamma\}$ denotes the relative orderings of elements come from
 $O\{\beta\}$ and $O\{\alpha\}$.

With the definition of momentum kernel \eqref{kernel}, we can write down the monodromy relations in string theory.
The general formula of KK-BCJ relation\cite{BjerrumBohr:2009rd} for color-ordered open string tree amplitudes
can be written as
\bea\label{KK_BCJ}
A_o(\beta_s,...,\beta_1,\alpha_1,...,\alpha_r,N)+(-1)^{s-1}\Sl_{\sigma\in P(O\{\alpha\}\bigcup O\{\beta\}),\sigma_1=\alpha_1}\mathcal{P}_{\{\beta^T,\alpha,N\},\{\sigma,N\}}A_o(\sigma,N)=0,
\eea
where we use $A_o$ to denote the color-ordered open string tree amplitudes and use $\beta^T$ to denote the revised
ordering of legs in permutation $\beta$. When taking the real part of the above equation, we get
the KK relations. The imaginary part of amplitudes must vanish. Thus, when taking the imaginary
part of the above equation, we get a set of constraints on the KK-basis. These constraints are
nothing but BCJ relations.  As in the case of pure KK relations, the KK-BCJ relations\eqref{KK_BCJ}
also express any amplitude by $(N-2)!$ amplitudes explicitly. However, in this case,
the coefficients of amplitudes are complex ones. The further reductions given by BCJ relations
 are implied by the vanishing of the imaginary parts of amplitudes.
If there is only one element in $\{\beta\}$, \eqref{KK_BCJ} becomes
 \bea\label{fun_u1_d}
&&A_o(\beta_1,\alpha_1,\alpha_2,...,\alpha_r,N)
+e^{-2i\pi\alpha'k_{\beta_1}\cdot k_{\alpha_1}}A_o(\alpha_1,\beta_1,\alpha_2,...,\alpha_r,N)\nn
&+&e^{-2i\pi\alpha'(k_{\beta_1}\cdot k_{\alpha_1}
+k_{\beta_1}\cdot k_{\alpha_2})}A_o(\alpha_1,\alpha_2,\beta_1,...,\alpha_r,N)\nn
&+&...+e^{-2i\pi\alpha'(k_{\beta_1}\cdot k_{\alpha_1}
+k_{\beta_1}\cdot k_{\alpha_2}
+...+k_{\beta_1}\cdot k_{\alpha_r})}A_o(\alpha_1,\alpha_2,...,\alpha_r,\beta_1,N)\nn
&=&0.
\eea
We call it  $U(1)$-like decoupling identity\footnote{We will see in the next subsection, this
is not a real $U(1)$-decoupling identity. However, with the new defined commutator in
the next subsection, this identity can be understood similarly as in the case of
$U(1)$-decoupling identity in field theory.}. The real part of this identity
is the fundamental KK relation(its field theory limit gives the $U(1)$-decoupling
identity in field theory)
 \bea
&&A_o(\beta_1,\alpha_1,\alpha_2,...,\alpha_r,N)+
\cos[2\pi\alpha'k_{\beta_1}\cdot k_{\alpha_1}]A_o(\alpha_1,\beta_1,\alpha_2,...,\alpha_r,N)\nn
&+&\cos[2\pi\alpha'(k_{\beta_1}\cdot k_{\alpha_1}
+k_{\beta_1}\cdot k_{\alpha_2})]A_o(\alpha_1,\alpha_2,\beta_1,...,\alpha_r,N)\nn
&+&...+\cos[2\pi\alpha'(k_{\beta_1}\cdot k_{\alpha_1}
+k_{\beta_1}\cdot k_{\alpha_2}
+...+k_{\beta_1}\cdot k_{\alpha_r})]A_o(\alpha_1,\alpha_2,...,\alpha_r,\beta_1,N)\nn
&=&0.\label{fun_KK}
\eea
The imaginary part of the $U(1)$-like decoupling identity \eqref{fun_u1_d} gives the
fundamental BCJ relation(its field theory limit gives fundamental BCJ relation in field theory)
\bea
&&\sin[2\pi\alpha'k_{\beta_1}\cdot k_{\alpha_1}]A_o(\alpha_1,\beta_1,\alpha_2,...,\alpha_r,N)
+\sin[2\pi\alpha'(k_{\beta_1}\cdot k_{\alpha_1}
+k_{\beta_1}\cdot k_{\alpha_2})]A_o(\alpha_1,\alpha_2,\beta_1,...,\alpha_r,N)\nn
&+&...+\sin[2\pi\alpha'(k_{\beta_1}\cdot k_{\alpha_1}
+k_{\beta_1}\cdot k_{\alpha_2}
+...+k_{\beta_1}\cdot k_{\alpha_r})]A_o(\alpha_1,\alpha_2,...,\alpha_r,\beta_1,N)\nn
&=&0.\label{fun_BCJ}
\eea

Another kind of useful monodromy relation that will be proven and used in the next
 section is the  generalized $U(1)$-like decoupling identity
\bea\label{Gen_u1_str}
\Sl_{\sigma\in P(O\{\beta_1,...,\beta_s\}\bigcup O\{\alpha_1,...,\alpha_r\})}
\mathcal{P}_{\{\beta,\alpha,N\},\{\sigma,N\}}A_o(\sigma,N)=0.
\eea
It is easy to see that the fundamental $U(1)$-like decoupling identity\eqref{fun_u1_d}
is the special case of the above equation with only one $\beta$.

Both the KK-BCJ relation and the generalized $U(1)$-like decoupling identity
can be considered as the special cases of the following general monodromy relation
\bea\label{generalized_KK_BCJ}
&&\Sl_{\tau\in P(O\{\gamma\}\bigcup O\{\beta^T\})}\mathcal{P}^*_{\{\gamma,\beta^T,\alpha,N\},\{\tau,\alpha,N\}}A_o(\tau,\alpha,N)\nn
&+&(-1)^{s-1}\Sl_{\sigma\in P(O\{\alpha\}\bigcup O\{\beta\})|\sigma_1=\alpha_1}\mathcal{P}_{\{\gamma,\beta^T,\alpha,N\},\{\gamma,\sigma,N\}}A_o(\gamma,\sigma,N)\nn
&=&0,
\eea
where the $\mathcal{P}^*$ is the complex conjugate of $\mathcal{P}$. In fact, if there is
no element in $\{\gamma\}$, the above equation becomes KK-BCJ relations\eqref{KK_BCJ}.
If there is no element in $\{\alpha\}$, the above equation becomes the complex conjugates
 of generalized $U(1)$-like decoupling identities\eqref{Gen_u1_str}. The general monodromy
 relations will be derived in section \ref{section_4}.

With this definition of momentum kernel,
we can also write down KLT relations(See Eq. (1.1) in \cite{Kawai:1985xq}) directly
\bea\label{KLT_relation}
M_c(1,2,...,N)=\Sl_{\sigma,\tau}A_o(\sigma)\mathcal{P}^*_{\{\sigma\},\{\tau\}}A_o(\tau),
\eea
where $M_c$ denote closed string tree amplitude. In the KLT relations \eqref{KLT_relation}
each open string takes half of the momentum of the corresponding closed string.

\subsection{The $(N-2)!$-formula of color decomposition in string theory}
\label{sub_color_decomposition}

In field theory, one can express the total $N$-gluon amplitudes at tree-level
 by either the standard color decomposition with $(N-1)!$ color-ordered amplitudes
  or the color decomposition with $(N-2)!$ color-ordered amplitudes(KK basis)
  \cite{Del Duca:1999ha}, \cite{DelDuca:1999rs}. This is because the $(N-1)!$
  amplitudes in the standard color decomposition can be expressed by $(N-2)!$
  KK basis. In string theory, when we introduce the gauge degree of freedom by
  adding Chan-Paton factors to the ends of open strings, the total amplitudes at
  tree-level for open strings can also be given by the standard color decomposition
\bea\label{standard_decomposition}
M_o(1^{a_1},...,N^{a_N})=\Sl_{\sigma\in S_{N-1}}Tr(T^{a_{\sigma_1}},...,T^{a_{\sigma_{N-1}}}T^{a_{N}})A_o(\sigma,N).
\eea
However, as we have mentioned, there are nontrivial kinematic factors before
the amplitudes in the KK relations in string theory. Thus the $(N-2)!$-formula
of color decomposition in string theory should be reconsidered. In this subsection,
we will show that the $(N-2)!$-formula of color decomposition in string theory
can be given as
\bea\label{new-C-O-D}
M_o(1,...,N)=\Sl_{\sigma\in S_{N-2}}Tr\left([[[T^{a_{1}},T^{a_{\sigma_2}}]_{\epsilon},T^{a_{\sigma_3}}]_{\epsilon}...
,T^{a_{\sigma_{N-1}}}]_{\epsilon}T^{a_{N}}\right) A_o(1,\sigma,N).
\eea
Here the commutator $[,]_{\epsilon}$ is defined as
\bea
[T^{a_1},T^{a_2}]_{\epsilon}=T^{a_1}T^{a_2}-e^{-2i\pi\alpha'k_1\cdot k_2}T^{a_2}T^{a_1}.
\eea
In this new defined commutator, there is a factor $e^{-2i\pi\alpha'k_1\cdot k_2}$
that depends on kinematic factor $k_1\cdot k_2$. The momentum of a commutator
$[T^{a_i},T^{a_j}]_{\epsilon}$ is defined as $k_i+k_j$.
After taking the field theory limits, we get the ordinary commutators in field theory
and the color decomposition becomes the one expressed by KK basis in field
theory\cite{DelDuca:1999rs}.

It is interesting that if there is some $T^a$ commute with all the other ones in
this new defined commutator, from the standard color decomposition
\eqref{standard_decomposition}, we obtain the relation
\eqref{fun_u1_d}. This is similar with the $U(1)$-decoupling identity in field theory,
but the commutators used are the new defined ones in which there are
  kinematic factors. That's why we  call \eqref{fun_u1_d} $U(1)$-like decoupling identity.
The generalized $U(1)$-like decoupling identities \eqref{Gen_u1_str} can be considered
as the extensions of $U(1)$-like decoupling identity \eqref{fun_u1_d}
\footnote{This is similar within the field theory case\cite{Du:2011js}.}. Before giving
the general discussion on this $(N-2)!$-formula of color decomposition, we first
give an example.

\subsubsection{Example}

We now take the four-point amplitude as an example. If \eqref{new-C-O-D} gives the right
color decomposition with $(N-2)!$ KK basis, we can transform \eqref{new-C-O-D} to the
standard color decomposition \eqref{standard_decomposition} by using KK-BCJ relation
\eqref{KK_BCJ}(which express the $(N-1)!$ amplitudes by $(N-2)!$ amplitudes)
\footnote{We can also use only KK relations(real part), however, in this paper,
we use the KK-BCJ relations. One thing should be further noticed is that
the imaginary parts of amplitudes must vanish.}.

The total four-point amplitude can be expressed by \eqref{new-C-O-D}
\bea\label{new-C-O-D-4}
M_o(1,2,3,4)
=Tr\left(\left[[T^{a_1},T^{a_2}]_{\epsilon},T^{a_3}\right]_{\epsilon}T^{a_4}\right)A_o(1,2,3,4)
+Tr\left(\left[[T^{a_1},T^{a_3}]_{\epsilon},T^{a_2}\right]_{\epsilon}T^{a_4}\right)A_o(1,3,2,4).
\eea
Expanding the new defined commutators in the traces, we get
\bea
&&M_o(1,2,3,4)\nn
&=&\Biggl[Tr(T^{a_1}T^{a_2}T^{a_3}T^{a_4})-e^{-2i\pi\alpha'k_2\cdot k_1}Tr(T^{a_2}T^{a_1}T^{a_3}T^{a_4})\nn
&&-e^{-2i\pi\alpha'(k_3\cdot k_1+k_3\cdot k_2)}Tr(T^{a_3}T^{a_1}T^{a_2}T^{a_4})+e^{-2i\pi\alpha'(k_3\cdot k_1+k_3\cdot k_2+k_2\cdot k_1)}Tr(T^{a_3}T^{a_2}T^{a_1}T^{a_4})\Biggr]A_o(1,2,3,4)\nn
&+&(2\leftrightarrow 3).
\eea
Now we collect the terms containing a same trace, e.g., for the trace $Tr(T^{a_1}T^{a_2}T^{a_3}T^{a_4})$,
we only have $A_o(1,2,3,4)$, while for
the trace $Tr(T^{a_2}T^{a_1}T^{a_3}T^{a_4})$, both the two permutations $1$, $2$, $3$, $4$ and
$1$, $3$, $2$, $4$ contribute to this trace because we have $O\{2,3\}\in P(O\{2\}\bigcup O\{3\})$ and
$O\{3,2\}\in P(O\{2\}\bigcup O\{3\})$,
we thus have
\bea
-e^{-2i\pi\alpha'k_2\cdot k_1}A_o(1,2,3,4)-e^{-2i\pi\alpha'(k_2\cdot k_1+k_2\cdot k_3)}A_o(1,3,2,4).
\eea
When we consider the KK-BCJ relation with $\beta_1=2$, $\alpha_1=3$, the above expression becomes
$A_o(2,1,3,4)$. In a similar way, for any given trace,
we can collect all the terms containing this trace and use an appropriate  KK-BCJ relation
to get an amplitude which has a same ordering within the trace. After considering all the traces, we get the
standard color decomposition
\bea
M_o(1,2,3,4)&=&Tr(T^{a_1}T^{a_2}T^{a_3}T^{a_4})A_o(1,2,3,4)+Tr(T^{a_1}T^{a_3}T^{a_2}T^{a_4})A_o(1,3,2,4)\nn
&+&Tr(T^{a_2}T^{a_1}T^{a_3}T^{a_4})A_o(2,1,3,4)+Tr(T^{a_2}T^{a_3}T^{a_1}T^{a_4})A_o(2,3,1,4)\nn
&+&Tr(T^{a_3}T^{a_1}T^{a_2}T^{a_4})A_o(3,1,2,4)+Tr(T^{a_3}T^{a_2}T^{a_1}T^{a_4})A_o(3,2,1,4).
\eea
Thus the color decomposition \eqref{new-C-O-D-4} is equivalent with the standard color decomposition.

\subsubsection{General discussion}

The discussion in the above example can be extended to the general proof.
In general, for a given $\sigma\in S_{N-2}$, we have a trace of the form
 \bea
 Tr\left([[[T^{a_{1}},T^{a_{\sigma_2}}]_{\epsilon},T^{a_{\sigma_3}}]_{\epsilon}...,T^{a_{\sigma_{N-1}}}]_{\epsilon}T^{a_{N}}\right).
 \eea
When we expand the commutators in this trace, we get $2^{N-2}$ terms. In each term, there is a trace of the form
\bea
Tr\left(T^{a_{\beta_s}}...T^{a_{\beta_1}}T^{a_1}T^{a_{\alpha_1}}...T^{a_{\alpha_{N-s-2}}}T^{a_N}\right).
\eea
This trace is accompanied by a phase factor when we consider the definition of momentum kernel \eqref{kernel}
\bea
(-1)^s\mathcal{P}_{\{\beta^T,1,\alpha,N\}\{1,\sigma,N\}}.
\eea
Any given $\sigma$ can split into two ordered sets $O\{\beta\}$ and $O\{\alpha\}$, where the relative orderings
of the legs in each ordered set are same with those in the
permutation $\sigma$. For a given $\sigma$, there are $C_{N-2}^0+C_{N-2}^1+...+C_{N-2}^{N-2}=2^{N-2}$
such splittings corresponding to the $2^{N-2}$ traces(we should notice that in the trace, the relative ordering
of $T^{a_{\beta_1}}$,...,$T^{a_{\beta_s}}$ is the reversed ordering of $\beta_1$,...,$\beta_s$ in $\sigma$).
Thus the total amplitude should be expressed as
\bea
M_o(1^{a_1},...,N^{a_N})&=&\Sl_{\sigma\in S_{N-2}}\Sl_{s=0}^{N-2}
\Sl_{\text{All splittings}\sigma\rightarrow O\{\alpha_1,...,\alpha_{N-s-2}\}\bigcup O\{\beta_s,...,\beta_1\}}(-1)^s\mathcal{P}_{\{\beta^T,1,\alpha,N\}\{1,\sigma,N\}}\nn
&&\times Tr\left(T^{a_{\beta_s}}...T^{a_{\beta_1}}T^{a_1}T^{a_{\alpha_1}}...T^{a_{\alpha_{N-s-2}}}T^{a_N}\right)
A(1,\sigma,N).
\eea
Because we sum over all permutations $\sigma\in S_{N-2}$, all the permutations
 $\sigma\in P(O\{\beta\}\bigcup O\{\alpha\})$
must be included. Thus, for a given trace
$Tr\left(T^{a_{\beta_s}}...T^{a_{\beta_1}}T^{a_1}T^{a_{\alpha_1}}
...T^{a_{\alpha_{N-s-2}}}T^{a_N}\right)$, we can collect all the permutations
which can split into two ordered sets $O\{\alpha\}$ and $O\{\beta\}$. We get
\bea
 (-1)^s\Sl_{\sigma\in P(O\{\alpha\}\bigcup O\{\beta\})}
 \mathcal{P}_{\{\beta^T,1,\alpha,N\},\{1,\sigma,N\}}A_o(1,\sigma,N)=A_o(\beta^T,1,\alpha,N),
 \eea
where we have used KK-BCJ relations. The ordering of the legs in $A_o(\beta^T,1,\alpha,N)$
is same with the ordering of the $T$s in the trace. After considering all the traces with different
$O\{\beta\}$s, we get the standard color
 decomposition \eqref{standard_decomposition}.

\section{Generating KK and BCJ relations by primary relations}\label{section_3}

As we have mentioned in the introduction, one can use the amplitude relations including the
cyclic symmetry, KK  and BCJ relations to reduce the number of the independent
$N$-point color-ordered amplitudes from $N!$ down to $(N-3)!$. It is interesting that all
the BCJ relations can be solved from a set of fundamental BCJ relations. In this section,
we will consider the question: When we consider all the cyclic symmetry, KK and BCJ relations
 together, can we generate all these relations by some primary relations? We will
 show that among all these relations, there are two primary relations.
 All other relations can be generated
  by the two primary relations. In string theory, one primary relation is the cyclic symmetry,
   while the other one can be chosen as either one of fundamental KK relation and fundamental
    BCJ relation. However, in field theory, if we do not consider the higher-order
     corrections, the fundamental KK relation(i.e., $U(1)$-decoupling identity in field
     theory) cannot be chosen as a primary relation. We can only choose the cyclic
     symmetry and the fundamental BCJ relation as the primary relations.
This is because the kinematic factors $s_{ij}$ which is necessary to generate BCJ relations
 do not appear in the $U(1)$-decoupling identity in field theory.

In the following subsections we will first show how to generate
$U(1)$-like decoupling identity \eqref{fun_u1_d} by two primary relations.
Then we will show the generalized $U(1)$-like decoupling identities \eqref{Gen_u1_str}
can be generated by
$U(1)$-like decoupling identities \eqref{fun_u1_d}. After that,
we will show the equivalence between generalized $U(1)$-like decoupling identities
\eqref{Gen_u1_str} and KK-BCJ relations \eqref{KK_BCJ}. Thus all the KK and BCJ relations
can be generated by the primary relations. We then will take the field theory limits
 to show how to generate all KK and BCJ relations in field theory.

\subsection{Generating $U(1)$-like decoupling identity by primary relations}

In this section, we will show if we know either the fundamental KK relation
or fundamental BCJ relation, we can generate the other one(then the whole $U(1)$-like
 decoupling identity) by using cyclic symmetry.  This means only two relations
  are the primary relations. The two primary relations can be chosen as the
cyclic symmetry
\bea\label{cyclic}
A_o(1,2,...,N)=A_o(N,1,...,N-1)
\eea
and arbitrary one of fundamental KK relation
 \eqref{fun_KK} and fundamental BCJ relation \eqref{fun_BCJ}.

Before our discussions, we should emphasize one thing. The $U(1)$-like decoupling
identity \eqref{fun_u1_d} can be understood as follows: We move $\beta_1$ from
the first location in the amplitude to the $(N-1)$-th location. When we move
$\beta_1$ from the left side to the right side of $\alpha_i$, we get a phase
factor $e^{-2i\pi\alpha'k_{\beta_1}\cdot k_{\alpha_i}}$ multiplied to the
corresponding amplitude. After consider all different locations of $\beta_1$,
we get the  $U(1)$-like decoupling identity  \eqref{fun_u1_d}. If the starting
point of $\beta_1$ is not the first location, e.g., the starting point is the
second location, we have another $U(1)$-like decoupling identity
\bea
&&A(\alpha_1,\beta_1,\alpha_2,...,\alpha_r,N)+e^{-2i\pi\alpha'k_{\beta_1}\cdot k_{\alpha_2}}A(\alpha_1,\alpha_2,\beta_1,...,\alpha_r,N)\nn
&&+...
+e^{-2i\pi\alpha'\left[k_{\beta_1}\cdot k_{\alpha_2}+...+k_{\beta_1}\cdot k_{\alpha_r}\right]}A(\alpha_1,...,\alpha_r,\beta_1,N)\nn
&&
+e^{-2i\pi\alpha'\left[k_{\beta_1}\cdot k_{\alpha_1}+...+k_{\beta_1}\cdot k_{\alpha_r}\right]}A(\alpha_1,...,\alpha_r,N,\beta_1)\nn
&&
=0.
\eea
This $U(1)$-like decoupling identity can be obtained by multiplying a
 phase factor $e^{-2i\pi\alpha'k_{\beta_1}\cdot k_{\alpha_1}}$ to both sides of \eqref{fun_u1_d}
  (Here the momentum conservation and on-shell condition should be considered).
  Thus the $U(1)$-like decoupling identities with different starting points of $\beta_1$
   are equivalent relations. If we know the $U(1)$-like decoupling identity with one
   starting point of $\beta_1$(e.g.,), we know all the $U(1)$-like decoupling identities
   (thus the fundamental KK and fundamental BCJ relations) with
    other starting points of $\beta_1$. In the following subsection, we will show that
    the $U(1)$-like decoupling identity \eqref{fun_u1_d} can be generated by two primary
    relations.

\subsubsection{Fundamental BCJ relation and cyclic symmetry as the primary relations}\label{subsection_BCJ_cyclic}

The fundamental KK relation \eqref{fun_KK} is not an independent relation
 if we consider the fundamental BCJ relation\eqref{fun_BCJ} and the cyclic symmetry\eqref{cyclic}
 as the primary ones. This is because any amplitude in the fundamental BCJ relation \eqref{fun_BCJ}
  has the cyclic symmetry \eqref{cyclic} and we can use the cyclic symmetry to change the starting point
of the leg $\beta_1$ from the second location to the third location to get another equivalent
 fundamental BCJ relation. Combining the two fundamental BCJ relations appropriately, we can derive
the real part condition(the fundamental KK relation) \eqref{fun_KK}. We now show the details.

When we consider the amplitudes with $\alpha_2$ as the first leg, according to the fundamental
 BCJ relation \eqref{fun_BCJ} with the starting point of $\beta_1$ at the second location,
we have
\bea
&&\sin(2\pi\alpha'k_{\beta_1}\cdot k_{\alpha_2})A_o(\alpha_2,\beta_1,\alpha_3,...,\alpha_r,N,\alpha_1)
+\sin[2\pi\alpha'(k_{\beta_1}\cdot k_{\alpha_2}
+k_{\beta_1}\cdot k_{\alpha_3})]A_o(\alpha_2,\alpha_3,\beta_1,...,\alpha_r,N,\alpha_1)\nn
&+&...+\sin[2\pi\alpha'(k_{\beta_1}\cdot k_{\alpha_2}
+k_{\beta_1}\cdot k_{\alpha_3}
+...+k_{\beta_1}\cdot k_{N})]A_o(\alpha_2,\alpha_3,...,\alpha_r,N,\beta_1,\alpha_1)\nn
&=&0.
\eea
With the cyclic symmetry \eqref{cyclic}, this relation turns to an equivalent relation with
the starting point of $\beta_1$ at the third location
 \bea\label{cy_fun_BCJ}
&&\sin(2\pi\alpha'k_{\beta_1}\cdot k_{\alpha_2})A_o(\alpha_1,\alpha_2,\beta_1,\alpha_3,...,\alpha_r,N)
+\sin[2\pi\alpha'(k_{\beta_1}\cdot k_{\alpha_2}
+k_{\beta_1}\cdot k_{\alpha_3})]A_o(\alpha_1,\alpha_2,\alpha_3,\beta_1,...,\alpha_r,N)\nn
&+&...+\sin[2\pi\alpha'(k_{\beta_1}\cdot k_{\alpha_2}
+k_{\beta_1}\cdot k_{\alpha_3}
+...+k_{\beta_1}\cdot k_{N})]A_o(\alpha_1,\alpha_2,\alpha_3,...,\alpha_r,N,\beta_1)\nn
&=&0.
\eea
Using $\sin(A+B)=\sin A\cos B+\cos A\sin B$,  the cyclic symmetry \eqref{cyclic},
the fundamental BCJ relation\eqref{fun_BCJ}, momentum conservation, the on-shell conditions
 $m_i^2=-k_i^2$ and $\alpha'm_i^2\in \mathbb{Z}$(e.g., in 26-dimensional bosonic string theory,
 for tachyon $\alpha'm_i^2=-1$ for massless vector $\alpha'm_i^2=0$, in 10-dimensional
 superstring theory, for massless vector $\alpha'm_i^2=0$), the relation \eqref{cy_fun_BCJ}
  can be rewritten as
\bea
&&A_o(\beta_1,\alpha_1,\alpha_2,...,\alpha_r,N)+\cos[2\pi\alpha'(k_{\beta_1}\cdot k_{\alpha_1}+k_{\beta_1}\cdot k_{\alpha_2})]A_o(\alpha_1,\alpha_2,\beta_1,...,\alpha_r,N)\nn
&+&...+\cos[2\pi\alpha'(k_{\beta_1}\cdot k_{\alpha_1}+k_{\beta_1}\cdot k_{\alpha_2}
+...+k_{\beta_1}\cdot k_{\alpha_r})]A_o(\alpha_1,\alpha_2,...,\alpha_r,\beta_1,N)\nn
&-&\cot(2\pi\alpha' k_{\beta_1}\cdot k_{\alpha_1})\Bigl\{\sin[2\pi\alpha'(k_{\beta_1}\cdot k_{\alpha_2}
+k_{\beta_1}\cdot k_{\alpha_1})]A_o(\alpha_1,\alpha_2,\beta_1,...,\alpha_r,N)\nn
&+&\sin[2\pi\alpha'(k_{\beta_1}\cdot k_{\alpha_2}
+k_{\beta_1}\cdot k_{\alpha_3}+k_{\beta_1}\cdot k_{\alpha_1})]A_o(\alpha_1,\alpha_2,\alpha_3,\beta_1,...,\alpha_r,N)\nn
&+&...+\sin[2\pi\alpha'(k_{\beta_1}\cdot k_{\alpha_2}
+k_{\beta_1}\cdot k_{\alpha_3}
+...+k_{\beta_1}\cdot k_{N}+k_{\beta_1}\cdot k_{\alpha_1})]A_o(\alpha_2,\alpha_3,...,\alpha_r,N,\beta_1,\alpha_1)\Bigr\}\nn
&=&A_o(\beta_1,\alpha_1,\alpha_2,...,\alpha_r,N)
+\cos(2\pi\alpha'k_{\beta_1}\cdot k_{\alpha_1})A_o(\alpha_1,\beta_1,\alpha_2,...,\alpha_r,N)\nn
&+&\cos[2\pi\alpha'(k_{\beta_1}\cdot k_{\alpha_1}+k_{\beta_1}\cdot k_{\alpha_2})]A_o(\alpha_1,\alpha_2,\beta_1,...,\alpha_r,N)\nn
&+&...+\cos[2\pi\alpha'(k_{\beta_1}\cdot k_{\alpha_1}+k_{\beta_1}\cdot k_{\alpha_2}
+...+k_{\beta_1}\cdot k_{\alpha_r})]A_o(\alpha_1,\alpha_2,...,\alpha_r,\beta_1,N)\nn
&=&0.
\eea
This just gives the fundamental KK relation  \eqref{fun_KK}. Thus we have derived the fundamental KK relation
by using the fundamental BCJ relation and cyclic symmetry as the primary relations.

\subsubsection{Fundamental KK relation and cyclic symmetry as the primary relations}

We can also consider the fundamental KK relation \eqref{fun_KK} and the cyclic symmetry \eqref{cyclic}
 as the primary relations. In this case, the fundamental BCJ relation \eqref{fun_BCJ} can be generated.
Actually, as what we have shown in the previous subsection, using the cyclic symmetry,
we can also move the starting point of $\beta_1$ in \eqref{fun_KK} from the first location
to the second location
\bea
&&A_o(\alpha_1,\beta_1,\alpha_2,\alpha_3,...,\alpha_r,N)+
\cos(2\pi\alpha'k_{\beta_1}\cdot k_{\alpha_2})A_o(\alpha_1,\alpha_2,\beta_1,\alpha_3,...,\alpha_r,N)\nn
&&+\cos[2\pi\alpha'(k_{\beta_1}\cdot k_{\alpha_2}
+k_{\beta_1}\cdot k_{\alpha_3})]A_o(\alpha_1,\alpha_2,\alpha_3,\beta_1,...,\alpha_r,N)\nn
&+&...+\cos[2\pi\alpha'(k_{\beta_1}\cdot k_{\alpha_2}
+k_{\beta_1}\cdot k_{\alpha_3}
+...+k_{\beta_1}\cdot k_{\alpha_r}+k_{\beta_1}\cdot k_N)]A_o(\alpha_1,\alpha_2,\alpha_3,...,\alpha_r,N,\beta_1)\nn
&=&0.\label{cy_fun_KK}
\eea
Using $\cos(A+B)=\cos A\cos B-\sin A\sin B$, the momentum conservation, the cyclic symmetry \eqref{cyclic}
 the on-shell conditions and the fundamental KK relation\eqref{fun_KK}, the above equation becomes
\bea
&&\cot(2\pi\alpha'k_{\beta_1}\cdot k_{\alpha_1})\Bigl\{A_o(\beta_1,\alpha_1,\alpha_2,...,\alpha_r,N)+\cos(2\pi\alpha'k_{\beta_1}\cdot k_{\alpha_1})A_o(\alpha_1,\beta_1,\alpha_2,\alpha_3,...,\alpha_r,N)\nn
&+&\cos[2\pi\alpha'(k_{\beta_1}\cdot k_{\alpha_1}+k_{\beta_1}\cdot k_{\alpha_2})]A_o(\alpha_1,\alpha_2,\beta_1,\alpha_3,...,\alpha_r,N)\nn
&+&...+\cos[2\pi\alpha'(k_{\beta_1}\cdot k_{\alpha_1}+k_{\beta_1}\cdot k_{\alpha_2}+...+k_{\beta_1}\cdot k_{\alpha_r})]A_o(\alpha_1,\alpha_2,\alpha_3,...,\alpha_r,\beta_1,N)\Bigr\}\nn
&+&\sin(2\pi\alpha'k_{\beta_1}\cdot k_{\alpha_1})A_o(\alpha_1,\beta_1,\alpha_2,...,\alpha_r,N)+\sin[2\pi\alpha'(k_{\beta_1}\cdot k_{\alpha_1}
+k_{\beta_1}\cdot k_{\alpha_2})]A_o(\alpha_1,\alpha_2,\beta_1,...,\alpha_r,N)\nn
&+&...+\sin[2\pi\alpha'(k_{\beta_1}\cdot k_{\alpha_1}
+k_{\beta_1}\cdot k_{\alpha_2}
+...+k_{\beta_1}\cdot k_{\alpha_r})]A_o(\alpha_1,\alpha_2,...,\alpha_r,\beta_1,N)\nn
&=&\sin(2\pi\alpha'k_{\beta_1}\cdot k_{\alpha_1})A_o(\alpha_1,\beta_1,\alpha_2,...,\alpha_r,N)+\sin[2\pi\alpha'(k_{\beta_1}\cdot k_{\alpha_1}
+k_{\beta_1}\cdot k_{\alpha_2})]A_o(\alpha_1,\alpha_2,\beta_1,...,\alpha_r,N)\nn
&+&...+\sin[2\pi\alpha'(k_{\beta_1}\cdot k_{\alpha_1}
+k_{\beta_1}\cdot k_{\alpha_2}
+...+k_{\beta_1}\cdot k_{\alpha_r})]A_o(\alpha_1,\alpha_2,...,\alpha_r,\beta_1,N)\nn
&=&0.
\eea
This is just the fundamental BCJ relation \eqref{fun_BCJ}.
So we have shown that the fundamental KK relation and the cyclic symmetry
can also be considered as the primary relations. With this choice of primary relations,
the fundamental BCJ relation can be generated.

\subsection{Generating generalized $U(1)$-like decoupling identity by $U(1)$-like decoupling identity}

In previous subsection, we have seen that $U(1)$-like decoupling identity can
be generated by two primary relations. In this subsection, we will show the generalized
$U(1)$-like decoupling identities \eqref{Gen_u1_str} with more $\beta$s is redundant relation,
i.e., they can be generated by the $U(1)$-like decoupling identity \eqref{fun_u1_d}.

To see this, we write the L. H. S. of \eqref{Gen_u1_str} as
\bea\label{lhs_gen_u1}
\mathcal{U}(\beta_1,...,\beta_s;\alpha_1,...,\alpha_r|N)\equiv\Sl_{\sigma\in P(O\{\beta_1,...,\beta_s\}\bigcup O\{\alpha_1,...,\alpha_r\})}\mathcal{P}_{\{\beta,\alpha,N\},\{\sigma,N\}}A_o(\sigma,N).
\eea
 To show \eqref{Gen_u1_str} with $s(s>1)$ $\beta$s(we mention it as level-$s$ relation)
 can be generated by the $U(1)$-like decoupling identity,
we will show $\mathcal{U}(\beta_1,...\beta_s;\alpha_1,...,\alpha_r|N)$ can be given
 as linear combinations of those $\mathcal{U}$s with $\beta$s fewer than $s$.
Then we get a recursive relation. The starting point of this recursion is
the $U(1)$-like decoupling identity \eqref{fun_u1_d}.
Once the $U(1)$-like decoupling identity \eqref{fun_u1_d} holds, the generalized
$U(1)$-like decoupling identities \eqref{Gen_u1_str} with more than one $\beta$s
 must also hold.

It will be convenience for us to introduce the following linear combinations
\bea\label{def_u}
\mathcal{U}(\gamma_1,...,\gamma_t;\beta_1,...,\beta_s;\alpha_1,...,\alpha_r|N)\equiv\Sl_{\tau\in P(O\{\alpha\}\bigcup O\{\beta\})}\mathcal{P}_{\{\gamma,\beta,\alpha,N\},\{\gamma,\tau,N\}}\mathcal{U}(\gamma_1,...,\gamma_t;\tau_1,...,\tau_{s+r}|N).
\eea
where $\mathcal{U}(\gamma_1,...,\gamma_t;\tau_1,...,\tau_{s+r}|N)$ is defined
by \eqref{lhs_gen_u1}.
$\mathcal{U}(\gamma_1,...,\gamma_t;\beta_1,...,\beta_s;\alpha_1,...,\alpha_r|N)$
has other two useful forms which can be obtained from \eqref{def_u}, \eqref{lhs_gen_u1}
and the factorization property of momentum kernels \eqref{kernel_5}. One is
\bea\label{prop_of_I1}
\mathcal{U}(\gamma_1,...,\gamma_t;\beta_1,...,\beta_s;\alpha_1,...,\alpha_r|N)=\Sl_{\sigma\in P(O\{\gamma\}\bigcup O\{\beta\}\bigcup O\{\alpha\})}\mathcal{P}_{\{\gamma,\beta,\alpha,N\},\{\sigma,N\}}A(\sigma,N).
\eea
The other one is
\bea\label{prop_of_I2}
\mathcal{U}(\gamma_1,...,\gamma_t;\beta_1,...,\beta_s;\alpha_1,...,\alpha_r|N)
=\Sl_{\tau\in P(O\{\gamma\}\bigcup O\{\beta\})}\mathcal{P}_{\{\gamma,\beta,\alpha,N\},\{\tau,\alpha,N\}}\mathcal{U}(\tau_1,...,\tau_{t+s};\alpha_1,...,\alpha_r|N).
\eea

It is worth to point that the following recursive relation will  be useful. If
\bea\label{recursion_1}
a_{i+1}=b_{i+1}-b_{i},
\eea
where $i$ can be any integer satisfying $i\geq 0$, we can express $b_n$ for any $n$ by $a_i$ and $b_0$
\bea\label{recursion_2}
b_n-b_0=\Sl_{i=1}^{n}a_i.
\eea
This recursion relation will be used again and again in this paper.

To make the recursive construction more clearly, we first give some examples.
\subsubsection{Examples}

We now give some warm-up examples.

{\bf Level-1}

The first example is the relation with only one $\beta$. This is nothing but the $U(1)$-like
decoupling identity
\bea
\mathcal{U}(\beta_1;\alpha_1,...,\alpha_r|N)=0.
\eea

{\bf Level-2}

 In the case with two $\beta$s, we consider the following three linear combinations
 \eqref{def_u} with $\beta_1$ and $\beta_2$ distributed
 into the ordered sets $O\{\gamma\}$ and $O\{\beta\}$ in \eqref{def_u}.
 The first one is
 \bea
 &&\mathcal{U}(\emptyset;\beta_2,\beta_1;\alpha_1,...,\alpha_r|N)\nn
 &=&\mathcal{P}_{\{\beta_2,\beta_1,\alpha_1,...,\alpha_r,N\},\{\beta_2,\beta_1,\alpha_1,...,\alpha_r,N\}}
 \mathcal{U}(\beta_2,\beta_1;\alpha_1,...,\alpha_r|N)=\mathcal{U}(\beta_2,\beta_1;\alpha_1,...,\alpha_r|N).
 \eea
 When we move $\beta_2$ into the ordered set $O\{\gamma\}$,
  we get the second one
 \bea
  \mathcal{U}(\beta_2;\beta_1;\alpha_1,...,\alpha_r|N)
  &=&\mathcal{P}_{\{\beta_2,\beta_1,\alpha,N\},\{\beta_2,\beta_1,\alpha,N\}}\mathcal{U}(\beta_2,\beta_1;\alpha_1,...,\alpha_r|N)\nn
  &+&\mathcal{P}_{\{\beta_2,\beta_1,\alpha,N\},\{\beta_1,\beta_2,\alpha,N\}}\mathcal{U}(\beta_1,\beta_2;\alpha_1,...,\alpha_r|N),
 \eea
 where we have used \eqref{prop_of_I2}.
 When we further move $\beta_1$ into the ordered set $O\{\gamma\}$, the ordered set $O\{\beta\}$ then becomes empty,
 while the $O\{\gamma\}$ set has two elements with the relative ordering $\beta_1$, $\beta_2$.
In this case we get the third one
 \bea
 &&\mathcal{U}(\beta_1,\beta_2;\emptyset;\alpha_1,...,\alpha_r|N)\nn
 &=&\mathcal{P}_{\{\beta_1,\beta_2,\alpha_1,...,\alpha_r,N\},\{\beta_1,\beta_2,\alpha_1,...,\alpha_r,N\}}\mathcal{U}(\beta_1,\beta_2;\alpha_1,...,\alpha_r|N)\nn
 &=&\mathcal{U}(\beta_2,\beta_1;\alpha_1,...,\alpha_r|N)\nn
 &=&e^{2i\pi\alpha'k_{\beta_1}\cdot k_{\beta_2}}\mathcal{P}_{\{\beta_2,\beta_1,\alpha_1,...,\alpha_r,N\},\{\beta_1,\beta_2,\alpha_1,...,\alpha_r,N\}}\mathcal{U}(\beta_1,\beta_2;\alpha_1,...,\alpha_r|N),
 \eea
 where the property of momentum kernel \eqref{kernel_3} has been used in the last line. If we define
 \bea
&& a_1=e^{-2i\pi\alpha'k_{\beta_1}\cdot k_{\beta_2}}\mathcal{U}(\beta_1,\beta_2;\emptyset;\alpha_1,...,\alpha_r|N), a_2=-\mathcal{U}(\beta_2;\beta_1;\alpha_1,...,\alpha_r|N),\nn
&&a_3=\mathcal{U}(\emptyset;\beta_2,\beta_1;\alpha_1,...,\alpha_r|N)\nn
& &b_0=0, b_1=\mathcal{P}_{\{\beta_2,\beta_1,\alpha,N\},\{\beta_1,\beta_2,\alpha,N\}}\mathcal{U}(\beta_1,\beta_2;\alpha_1,...,\alpha_r|N),\nn
&&b_2=-\mathcal{P}_{\{\beta_2,\beta_1,\alpha_1,...,\alpha_r,N\},\{\beta_2,\beta_1,\alpha_1,...,\alpha_r,N\}}
\mathcal{U}(\beta_2,\beta_1;\alpha_1,...,\alpha_r|N),
 b_3=0,
 \eea
 the relation \eqref{recursion_1} is satisfied for $i=0,1,2$. Thus  \eqref{recursion_2} gives
 \bea
 e^{-2i\pi\alpha'k_{\beta_1}\cdot k_{\beta_2}}\mathcal{U}(\beta_1,\beta_2;\alpha_1,...,\alpha_r|N)
 -\mathcal{U}(\beta_2;\beta_1;\alpha_1,...,\alpha_r|N)+\mathcal{U}(\beta_2,\beta_1;\alpha_1,...,\alpha_r|N)=0.
 \eea
 Exchanging $\beta_1$ and $\beta_2$, we have another equation
 \bea
 e^{-2i\pi\alpha'k_{\beta_2}\cdot k_{\beta_1}}\mathcal{U}(\beta_2,\beta_1;\alpha_1,...,\alpha_r|N)-\mathcal{U}(\beta_1;\beta_2;\alpha_1,...,\alpha_r|N)+\mathcal{U}(\beta_1,\beta_2;\alpha_1,...,\alpha_r|N)=0.
 \eea
 From the two equations above, we get
 \bea\label{level_2_U}
 \mathcal{U}(\beta_1,\beta_2;\alpha_1,...,\alpha_r|N)=\frac{1}{2i\sin(2\pi\alpha'k_{\beta_1}\cdot k_{\beta_2})}\left[e^{2i\pi\alpha'k_{\beta_1}\cdot k_{\beta_2}}\mathcal{U}(\beta_1;\beta_2;\alpha_1,...,\alpha_r|N)-\mathcal{U}(\beta_2;\beta_1;\alpha_1,...,\alpha_r|N)\right].\nn
 \eea
Using the definition of $\mathcal{U}(\gamma;\beta;\alpha|N)$ \eqref{def_u}, $\mathcal{U}(\beta_1;\beta_2;\alpha_1,...,\alpha_r|N)$ can be expressed as
\bea\label{2_by_1_U1}
\mathcal{U}(\beta_1;\beta_2;\alpha_1,...,\alpha_r|N)
=\Sl_{\tau\in P(O\{\alpha_1,...,\alpha_r\}\bigcup\{\beta_2\})}
\mathcal{P}_{\{\beta_1,\beta_2,\alpha_1,...,\alpha_r,N\},\{\beta_1,\tau_1,...,\tau_{1+r},N\}}\mathcal{U}(\beta_1;\tau_1,...,\tau_{r+1}|N),
\eea
while $\mathcal{U}(\beta_2;\beta_1;\alpha_1,...,\alpha_r|N)$ can be expressed as
\bea\label{2_by_1_U2}
\mathcal{U}(\beta_2;\beta_1;\alpha_1,...,\alpha_r|N)
=\Sl_{\tau\in P(O\{\alpha_1,...,\alpha_r\}\bigcup\{\beta_1\})}
\mathcal{P}_{\{\beta_2,\beta_1,\alpha_1,...,\alpha_r,N\},\{\beta_2,\tau_1,...,\tau_{1+r},N\}}\mathcal{U}(\beta_2;\tau_1,...,\tau_{r+1}|N).
\eea
Once we have $U(1)$-like decoupling identity, $\mathcal{U}(\beta_1;\beta_2;\alpha_1,...,\alpha_r|N)$
and $\mathcal{U}(\beta_2;\beta_1;\alpha_1,...,\alpha_r|N)$ must vanish,
then $\mathcal{U}(\beta_1,\beta_2;\alpha_1,...,\alpha_r|N)$ also vanishes.
Hence all the level-2 generalized $U(1)$-like decoupling identities are generated by
the $U(1)$-like decoupling identity \eqref{fun_u1_d}.

{\bf Level-3}

Now let us consider the level-3 relation. In this case, we should distribute
$\beta_1$ $\beta_2$, $\beta_3$ into the two ordered sets $O\{\gamma\}$ and
$O\{\beta\}$ in \eqref{def_u}.
We just use those combinations with keeping the relative orderings in the permutation $\beta_3$, $\beta_2$, $\beta_1$ in each ordered set, i.e.,
$\mathcal{U}(\emptyset;\beta_3,\beta_2,\beta_1;\alpha_1,...,\alpha_r|N)$, $\mathcal{U}(\beta_3;\beta_2,\beta_1;\alpha_1,...,\alpha_r|N)$, $\mathcal{U}(\beta_2,\beta_3;\beta_1;\alpha_1,...,\alpha_r|N)$ and  $\mathcal{U}(\beta_1,\beta_2,\beta_3;\emptyset;\alpha_1,...,\alpha_r|N)$.
Each one can be given by the expression \eqref{prop_of_I2}. In \eqref{prop_of_I2}, the first element in $\tau$$(\tau\in P(O\{\gamma\}\bigcup O\{\beta\}))$ can be either the first one in $\gamma$ or the
first one in $\beta$. Thus we can rewrite these combinations of amplitudes as
\bea\label{level-3-1}
&&\mathcal{U}(\emptyset;\beta_3,\beta_2,\beta_1;\alpha_1,...,\alpha_r|N)\nn
&=&\mathcal{P}_{\{\beta_3,\beta_2,\beta_1,\alpha_1,...,\alpha_r,N\},\{\beta_3,\beta_2,\beta_1,\alpha_1,...,\alpha_r,N\}}\mathcal{U}(\beta_3,\beta_2,\beta_1;\alpha_1,...,\alpha_r|N)\nn
&=&\mathcal{U}(\beta_3,\beta_2,\beta_1;\alpha_1,...,\alpha_r|N),
\eea
\bea\label{level-3-2}
&&\mathcal{U}(\beta_3;\beta_2,\beta_1;\alpha_1,...,\alpha_r|N)\nn
&=&\Sl_{\tau\in P(O\{\beta_3\}\bigcup O\{\beta_2,\beta_1\}),\tau_1=\beta_3}
\mathcal{P}_{\{\beta_3,\beta_2,\beta_1,\alpha_1,...,\alpha_r,N\},\{\tau,\alpha_1,...,\alpha_r,N\}}\mathcal{U}(\tau;\alpha_1,...,\alpha_r|N)\nn
&+&\Sl_{\tau\in P(O\{\beta_3\}\bigcup O\{\beta_2,\beta_1\}),\tau_1=\beta_2}
\mathcal{P}_{\{\beta_3,\beta_2,\beta_1,\alpha_1,...,\alpha_r,N\},\{\tau,\alpha_1,...,\alpha_r,N\}}\mathcal{U}(\tau;\alpha_1,...,\alpha_r|N).
\eea
\bea\label{level-3-3}
&&\mathcal{U}(\beta_2,\beta_3;\beta_1;\alpha_1,...,\alpha_r|N)\nn
&=&\Sl_{\tau\in P(O\{\beta_2,\beta_3\}\bigcup O\{\beta_1\}),\tau_1=\beta_2}\mathcal{P}_{\{\beta_2,\beta_3,\beta_1,\alpha_1,...,\alpha_r,N\},\{\tau,\alpha_1,...,\alpha_r,N\}}\mathcal{U}(\tau;\alpha_1,...,\alpha_r|N)\nn
&+&\Sl_{\tau\in P(O\{\beta_2,\beta_3\}\bigcup O\{\beta_1\}),\tau_1=\beta_1}\mathcal{P}_{\{\beta_2,\beta_3,\beta_1,\alpha_1,...,\alpha_r,N\},\{\tau,\alpha_1,...,\alpha_r,N\}}\mathcal{U}(\tau;\alpha_1,...,\alpha_r|N).
\eea
\bea\label{level-3-4}
&&\mathcal{U}(\beta_1,\beta_2,\beta_3;\emptyset;\alpha_1,...,\alpha_r|N)\nn
&=&\mathcal{P}_{\{\beta_1,\beta_2,\beta_3,\alpha_1,...,\alpha_r,N\},\{\beta_1,\beta_2,\beta_3,\alpha_1,...,\alpha_r,N\}}\mathcal{U}(\beta_1,\beta_2,\beta_3;\alpha_1,...,\alpha_r|N)\nn
&=&\mathcal{U}(\beta_1,\beta_2,\beta_3;\alpha_1,...,\alpha_r|N).
\eea
We should notice that the $\tau$ in \eqref{level-3-4} is same with that in the second term of \eqref{level-3-3}, the $\tau$ in the first term of \eqref{level-3-3} is
same with that in the second term of \eqref{level-3-2} while the $\tau$ in the first term of \eqref{level-3-2} is same with that in \eqref{level-3-1}.
Using  \eqref{def_u} to adjust the coefficients in each equation, we can define
\bea
&&a_4=\mathcal{U}(\emptyset;\beta_3,\beta_2,\beta_1;\alpha_1,...,\alpha_r|N),b_4=0,\nn
&&a_3=-\mathcal{U}(\beta_3;\beta_2,\beta_1;\alpha_1,...,\alpha_r|N), b_3=-\mathcal{P}_{\{\beta_3,\beta_2,\beta_1,\alpha_1,...,\alpha_r,N\},\{\beta_3,\beta_2,\beta_1,\alpha_1,...,\alpha_r,N\}},\nn
&&a_2=e^{-2i\pi\alpha'k_{\beta_2}\cdot k_{\beta_3}}\mathcal{U}(\beta_2,\beta_3;\beta_1 ;\alpha_1,...,\alpha_r|N) ,b_2=\mathcal{P}_{\{\beta_3,\beta_2,\beta_1,\alpha_1,...,\alpha_r,N\},\{\tau,\alpha_1,...,\alpha_r,N\}}\mathcal{U}(\tau;\alpha_1,...,\alpha_r|N),\nn
&&a_1=-e^{\left[-2i\pi\alpha'\left(k_{\beta_1}\cdot k_{\beta_2}+k_{\beta_2}\cdot k_{\beta_3}+k_{\beta_3}\cdot k_{\beta_1}\right)\right]}\mathcal{U}(\beta_1,\beta_2,\beta_3;\emptyset;\alpha_1,...,\alpha_r|N),\nn
&&b_1=-\mathcal{P}_{\{\beta_3,\beta_2,\beta_1,\alpha_1,...,\alpha_r,N\},\{\beta_1,\beta_2,\beta_3,\alpha_1,...,\alpha_r,N\}}\mathcal{U}(\tau;\alpha_1,...,\alpha_r|N),b_0=0.
\eea
These $a$s and $b$s satisfy \eqref{recursion_1}, thus they also satisfy \eqref{recursion_2}. We have
\bea
&&a_1+a_2+a_3+a_4\nn
&=&-e^{\left[-2i\pi\alpha'\left(k_{\beta_1}\cdot k_{\beta_2}+k_{\beta_2}\cdot k_{\beta_3}+k_{\beta_3}\cdot k_{\beta_1}\right)\right]}\mathcal{U}(\beta_1,\beta_2,\beta_3; \alpha_1,...,\alpha_r|N)\nn
&+&e^{-2i\pi\alpha'k_{\beta_2}\cdot k_{\beta_3}}\mathcal{U}(\beta_2,\beta_3;\beta_1 ;\alpha_1,...,\alpha_r|N)-\mathcal{U}(\beta_3;\beta_2,\beta_1;\alpha_1,...,\alpha_r|N)+\mathcal{U}(\beta_3,\beta_2,\beta_1;\alpha_1,...,\alpha_r|N)\nn
&=&0.
\eea
Now we replace  $\beta_1$, $\beta_2$, $\beta_3$ by $\beta_3$, $\beta_2$, $\beta_1$, i.e., reverse the relative ordering of legs in permutation  $\beta_1$, $\beta_2$, $\beta_3$,
we get another equation
\bea
&&-e^{\left[-2i\pi\alpha'\left(k_{\beta_1}\cdot k_{\beta_2}+k_{\beta_2}\cdot k_{\beta_3}+k_{\beta_3}\cdot k_{\beta_1}\right)\right]}\mathcal{U}(\beta_3,\beta_2,\beta_1; \alpha_1,...,\alpha_r|N)\nn
&+&e^{-2i\pi\alpha'k_{\beta_2}\cdot k_{\beta_1}}\mathcal{U}(\beta_2,\beta_1; \beta_3;\alpha_1,...,\alpha_r|N)-\mathcal{U}(\beta_1;\beta_2,\beta_3;\alpha_1,...,\alpha_r|N)+\mathcal{U}(\beta_1,\beta_2,\beta_3;\alpha_1,...,\alpha_r|N)\nn
&=&0.
\eea
From this two equations above, we can solve $\mathcal{U}(\beta_1,\beta_2,\beta_3; \alpha_1,...,\alpha_r|N)$ out
\bea
&&\mathcal{U}(\beta_1,\beta_2,\beta_3;\alpha_1,...,\alpha_r|N)\nn
&=&\frac{1}{2i\sin\left[2\pi\alpha'(k_{\beta_1}\cdot k_{\beta_2}+k_{\beta_2}\cdot k_{\beta_3}+k_{\beta_3}\cdot k_{\beta_1})\right]}\nn
&&\times\Biggl\{-e^{2i\pi(\alpha'k_{\beta_1}\cdot k_{\beta_3}+k_{\beta_2}\cdot k_{\beta_3})}\mathcal{U}(\beta_2,\beta_1;\beta_3;\alpha_1,...,\alpha_r|N)+\mathcal{U}(\beta_3;\beta_2,\beta_1;\alpha_1,...,\alpha_r,N)\nn
&-&e^{-ik_{\beta_2}\cdot k_{\beta_3}}\mathcal{U}(\beta_2,\beta_3;\beta_1;\alpha_1,...,\alpha_r,N)+e^{\left[2i\pi\alpha'\left(k_{\beta_1}\cdot k_{\beta_2}+k_{\beta_2}\cdot k_{\beta_3}+k_{\beta_3}\cdot k_{\beta_1}\right)\right]}\mathcal{U}(\beta_1;\beta_2,\beta_3;\alpha_1,...,\alpha_r,N)\Biggr\}.
\eea
Substituting all the  $\mathcal{U}(\gamma;\beta;\alpha|N)$s in the above equation by the definition \eqref{def_u}, as in case of level-2, we express $\mathcal{U}(\beta_1,\beta_2,\beta_3;\alpha_1,...,\alpha_r|N)$ by those $\mathcal{U}(\beta;\alpha|N)$s at level-2 and level-1.
Since the $U(1)$-like decoupling identities at level-2 have been generated by the $U(1)$-like decoupling identity,
the level-3 identities can also be generated by the $U(1)$-like decoupling identity.

\subsubsection{General discussion}

The discussions on level-2 and level-3 can be extended to the general case. In general,
we can distribute $\beta_1,...,\beta_s$ into two ordered sets
$O\{\beta_t,...,\beta_1\}$ and $O\{\beta_{t+1},...,\beta_s\}$ for any $t$. We consider
this two ordered sets as the sets $O\{\gamma\}$ and $O\{\beta\}$ in $\mathcal{U}(\gamma;\beta;\alpha|N)$.
 Then for a given $t$, we have $\mathcal{U}(\beta_{t+1},...,\beta_s;\beta_t,...,\beta_1;\alpha_1,...,\alpha_r|N)$
 which can be expressed  by \eqref{prop_of_I2}. Noticing the
 first element in $\tau$ can be either $\beta_{t+1}$ or $\beta_t$, we have
 \bea
&&\mathcal{U}(\beta_{t+1},...,\beta_s;\beta_t,...,\beta_1;\alpha_1,...,\alpha_r|N)\nn
&=&\Sl_{\tau\in P(O\{\beta_{t+1},...,\beta_s\}\bigcup O\{\beta_t,...,\beta_1\}),\tau_1=\beta_{t+1}}\mathcal{P}_{\{\beta_{t+1},...,\beta_s,\beta_t,...,\beta_1,\alpha_1,...,\alpha_r,N\},
\{\tau_1,...,\tau_{s},\alpha_1,...,\alpha_r,N\}}\mathcal{U}(\tau_1,...,\tau_s;\alpha_1,...,\alpha_r|N)\nn
&+&\Sl_{\tau\in P(O\{\beta_{t+1},...,\beta_s\}\bigcup O\{\beta_t,...,\beta_1\}),\tau_1=\beta_{t}}\mathcal{P}_{\{\beta_{t+1},...,\beta_s,\beta_t,...,\beta_1,\alpha_1,...,\alpha_r,N\},
\{\tau_1,...,\tau_{s},\alpha_1,...,\alpha_r,N\}}\mathcal{U}(\tau_1,...,\tau_s;\alpha_1,...,\alpha_r|N).\nn
\eea
For any given $t$ and a permutations $\tau\in P(O\{\beta_{t+1},...,\beta_s\}\bigcup O\{\beta_t,...,\beta_1\})(\tau_1=\beta_{t})$
there exists a $\tau'$ identical to $\tau$,
$\tau'\in P(O\{\beta_{t},...,\beta_s\}\bigcup O\{\beta_{t-1},...,\beta_1\})(\tau'_1=\beta_{t})$.
Thus, for a given $t$, the second term of
$\mathcal{U}(\beta_{t+1},...,\beta_s;\beta_t,...,\beta_1;\alpha_1,...,\alpha_r|N)$ is proportional
to the first term of $\mathcal{U}(\beta_{t},...,\beta_s;\beta_{t-1},...,\beta_1;\alpha_1,...,\alpha_r|N)$.
Adjusting the coefficients and using the property \eqref{kernel_3}, we can define
\bea
a_{t+1}&=&(-1)^{s-t}e^{-2i\pi\Sl_{t+1\leq i<j\leq s}\alpha'k_i\cdot k_j}\mathcal{U}(\beta_{t+1},...,\beta_s;\beta_t,...,\beta_1;\alpha_1,...,\alpha_r|N),(1\leq t \leq s-1)\nn
b_{t+1}&=&(-1)^{s-t}\Sl_{\tau\in P(O\{\beta_{t+1},...,\beta_s\}\bigcup O\{\beta_t,...,\beta_1\}),\tau_1=\beta_{t+1}}\mathcal{P}_{\{\beta_{s},...,\beta_1,\alpha_1,...,\alpha_r,N\},
\{\tau_1,...,\tau_{s},\alpha_1,...,\alpha_r,N\}}\nn
&&\times\mathcal{U}(\tau_1,...,\tau_s;\alpha_1,...,\alpha_r|N),(1\leq t \leq s-1),\nn
a_{s+1}&=&\mathcal{U}(\beta_s,...,\beta_1;\alpha_1,...,\alpha_r|N), b_{s+1}=0,\nn
a_1&=&(-1)^se^{-2i\pi\Sl_{1\leq i<j\leq s}\alpha'k_i\cdot k_j}\mathcal{U}(\beta_1,...,\beta_s;\alpha_1,...,\alpha_r|N),\nn
 b_1&=&(-1)^s\mathcal{P}_{\{\beta_{s},...,\beta_1,\alpha_1,...,\alpha_r,N\},
\{\beta_{1},...,\beta_s,\alpha_1,...,\alpha_r,N\}}\mathcal{U}(\beta_{1},...,\beta_s;\alpha_1,...,\alpha_r|N)\nn
b_0&=&0.
\eea
These definitions satisfy \eqref{recursion_1}, then from \eqref{recursion_2} we have
\bea
\Sl_{t=1}^{s+1}a_t
&=&\mathcal{U}(\beta_s,...,\beta_1;\alpha_1,...,\alpha_r|N)+(-1)^se^{-2i\pi\alpha'\Sl_{1\leq i<j\leq s}k_i\cdot k_j}\mathcal{U}(\beta_1,...,\beta_s;\alpha_1,...,\alpha_r|N)\nn
&&+\Sl_{t=1}^{s-1}(-1)^{s-t}e^{-2i\pi\alpha'\Sl_{t+1\leq i<j\leq s}k_i\cdot k_j}\mathcal{U}(\beta_{t+1},...,\beta_s;\beta_t,...,\beta_1;\alpha_1,...,\alpha_r|N)\nn
&=&0.
\eea
Now we reverse the ordering of $\beta_1,...,\beta_s$, i.e., we do the replacement
$\beta_1,...,\beta_s\rightarrow \beta_s,...,\beta_1$, we get another equation
\bea
&&\mathcal{U}(\beta_1,...,\beta_s;\alpha_1,...,\alpha_r|N)+(-1)^se^{-2i\pi\alpha'\Sl_{1\leq i<j\leq s}k_i\cdot k_j}\mathcal{U}(\beta_s,...,\beta_1;\alpha_1,...,\alpha_r|N)\nn
&&+\Sl_{t=1}^{s-1}(-1)^{t}e^{-2i\pi\alpha'\Sl_{1\leq i<j\leq t}k_i\cdot k_j}\mathcal{U}(\beta_t,...,\beta_1;\beta_{t+1},...,\beta_s;\alpha_1,...,\alpha_r|N)\nn
&=&0.
\eea
 $\mathcal{U}(\beta_1,...,\beta_s;\alpha_1,...,\alpha_r|N)$ can be solved  out from the above two equations
\bea
&&\mathcal{U}(\beta_1,...,\beta_s;\alpha_1,...,\alpha_r|N)\nn
&=&\frac{1}{2i\sin\left[2\pi\alpha'\Sl_{1\leq i<j\leq t}k_i\cdot k_j\right]}\nn
&&\times\Biggl[\Sl_{t=1}^{s-1}(-1)^{t}e^{-2i\pi\alpha'\Sl_{t+1\leq i<j\leq s}k_i\cdot k_j}\mathcal{U}(\beta_{t+1},...,\beta_s;\beta_t,...,\beta_1;\alpha_1,...,\alpha_r|N)\nn
&&-e^{2i\pi\alpha'\Sl_{t+1\leq i<j\leq s}k_i\cdot k_j}\Sl_{t=1}^{s-1}(-1)^{t}e^{-2i\pi\alpha'\Sl_{1\leq i<j\leq t}k_i\cdot k_j}\mathcal{U}(\beta_t,...,\beta_1;\beta_{t+1},...,\beta_s;\alpha_1,...,\alpha_r|N)\Biggr].
\eea
From \eqref{def_u} we know $\mathcal{U}(\beta_{t+1},...,\beta_s;\beta_t,...,\beta_1;\alpha_1,...,\alpha_r|N)$ and
$\mathcal{U}(\beta_t,...,\beta_1;\beta_{t+1},...,\beta_s;\alpha_1,...,\alpha_r|N)$ are defined by combinations of $\mathcal{U}(\beta;\alpha;N)$s with $\beta$s fewer than $s$.
If the generalized $U(1)$-like decoupling identities at level-$i(i<s)$ can be generated
by the $U(1)$-like decoupling identity,  the generalized $U(1)$-like decoupling identity at level-$s$ can also be generated by
the $U(1)$-like decoupling identity.

\subsection{Generating KK-BCJ relation by generalized $U(1)$-like decoupling identity}

So far, we have shown that all the generalized $U(1)$-like decoupling identities can be generated
by $U(1)$-like decoupling identity. In this subsection, let us turn to the KK-BCJ relation\eqref{KK_BCJ}.
We will show the generalized $U(1)$-like decoupling identities and the KK-BCJ relations can be solved from each other,
i.e., they are equivalent relations. Thus all the KK-BCJ relations
are generated by the primary relations. Since the KK and BCJ relations are just the real part and imaginary part
 of KK-BCJ relations, they can also be generated by primary relations.
To show the equivalence of KK-BCJ relations and generalized $U(1)$-like decoupling identities,
we define another useful linear combination of amplitudes by the L. H. S. of KK-BCJ relations \eqref{KK_BCJ}
\bea\label{lhs_KK_BCJ}
& &\mathcal{V}(\beta_1,...,\beta_s;\alpha_1,...,\alpha_r|N)\nn
&\equiv& A_o(\beta_s,...,\beta_1,\alpha_1,...,\alpha_r,N)
+(-1)^{s-1}\Sl_{\sigma\in P(O\{\alpha\}\bigcup O\{\beta\}),\sigma_1=\alpha_1}\mathcal{P}_{\{\beta^T,\alpha,N\},\{\sigma,N\}}A_o(\sigma,N).\nn
\eea
The equivalence between KK-BCJ relation and generalized $U(1)$-like decoupling identity means
$\mathcal{V}(\beta;\alpha|N)$ and $\mathcal{U}(\beta_;\alpha|N)$ can be expressed by each other.
We first give some examples.
\subsubsection{Examples}

In this subsection, we give some examples to show the equivalence between generalized $U(1)$-like
decoupling identities and the KK-BCJ relations.

{\bf Level-1}

In section \ref{section_2}, we have already mentioned that the KK-BCJ relation is same with the
generalized $U(1)$-like decoupling identity at level-1. At this level, they all become
 $U(1)$-like decoupling identity \eqref{fun_u1_d}, i.e.,
\bea
\mathcal{U}(\beta_1;\alpha_1,...,\alpha_r|N)=\mathcal{V}(\beta_1;\alpha_1,...,\alpha_r|N)=0.
\eea

{\bf Level-2}

Now we consider level-2. The difference between $\mathcal{V}(\beta_1,\beta_2;\alpha|N)$ at level-2
and $\mathcal{V}(\beta_2;\beta_1,\alpha|N)$ at level-1 is
\bea
&&\mathcal{V}(\beta_1,\beta_2;\alpha_1,...,\alpha_r|N)-\mathcal{V}(\beta_2;\beta_1,\alpha_1,...,\alpha_r|N)\nn
&=&A_o(\beta_2,\beta_1,\alpha_1,...,\alpha_r,N)+(-1)\Sl_{\sigma\in P(O\{\alpha_1,...,\alpha_r\}\bigcup O\{\beta_1,\beta_2\}),\sigma_1=\alpha_1}\mathcal{P}_{\{\beta_2,\beta_1,\alpha,N\},\{\sigma,N\}}A_o(\sigma,N)\nn
&&-\left[A_o(\beta_2,\beta_1,\alpha_1,...,\alpha_r,N)+\Sl_{\sigma'\in P(O\{\beta_1,\alpha_1,...,
\alpha_r\}\bigcup O\{\beta_2\}),\sigma_1=\beta_1}\mathcal{P}_{\{\beta_2,\beta_1,\alpha,N\},\{\sigma',N\}}A_o(\sigma',N)\right]\nn
&=&-\Sl_{\sigma\in P(O\{\alpha_1,...,\alpha_r\}\bigcup O\{\beta_1,\beta_2\})}\mathcal{P}_{\{\beta_2,\beta_1,\alpha,N\},\{\sigma,N\}}A_o(\sigma,N)\nn
&=&-e^{-2i\pi\alpha'k_{\beta_1}\cdot k_{\beta_2}}\Sl_{\sigma\in P(O\{\alpha_1,...,\alpha_r\}\bigcup O\{\beta_1,\beta_2\})}
\mathcal{P}_{\{\beta_1,\beta_2,\alpha,N\},\{\sigma,N\}}A_o(\sigma,N),
\eea
where we have used the property \eqref{kernel_3}. The last line in the above equation is just
$\mathcal{U}(\beta_1,\beta_2;\alpha_1,...,\alpha_r|N)$ multiplied by a factor $-e^{-2i\pi\alpha'k_1\cdot k_2}$.
Thus we can express the $\mathcal{U}(\beta_1,\beta_2;\alpha_1,...,\alpha_r|N)$ at
level-2 by those $\mathcal{V}$s at level-2 and level-1
\bea
\mathcal{U}(\beta_1,\beta_2;\alpha_1,...,\alpha_r|N)
=-e^{2i\pi\alpha'k_{\beta_1}\cdot k_{\beta_2}}\left[\mathcal{V}(\beta_1,\beta_2;\alpha_1,...,\alpha_r|N)-\mathcal{V}(\beta_2;\beta_1,\alpha_1,...,\alpha_r|N)\right].
\eea
If we define
\bea
&&a_2=-e^{-2i\pi\alpha'k_{\beta_1}\cdot k_{\beta_2}}\mathcal{U}(\beta_1,\beta_2;\alpha_1,...,\alpha_r|N),b_2=\mathcal{V}(\beta_1,\beta_2;\alpha_1,...,\alpha_r|N),\nn
&&a_1=\mathcal{U}(\beta_2;\beta_1,\alpha_1,...,\alpha_r|N), b_1=\mathcal{V}(\beta_2;\beta_1,\alpha_1,...,\alpha_r|N),\nn
&&b_0=0,
\eea
the relation \eqref{recursion_1} is satisfied. Thus we can use \eqref{recursion_2} to express $b_2$ as
\bea\label{level_2_V}
\mathcal{V}(\beta_1,\beta_2;\alpha_1,...,\alpha_r|N)
=\mathcal{U}(\beta_2;\beta_1,\alpha_1,...,\alpha_r|N)-e^{-2i\pi\alpha'k_{\beta_1}\cdot k_{\beta_2}}\mathcal{U}(\beta_1,\beta_2;\alpha_1,...,\alpha_r|N).
\eea
This means we can also use the generalized $U(1)$-like decoupling identities
 at level-2 and level-1 to express the KK-BCJ relation at level-2. Thus the KK-BCJ
relations at level-$s$($s\leq 2$)  are equivalent with the generalized
$U(1)$-like decoupling identity at level-$s$($s\leq2$).
Since we have shown that all
the generalized $U(1)$-like decoupling identities can be generated by the
 primary relations, the KK-BCJ relation at level-2 can  also be generated
 by the primary relations.

{\bf level-3}

Now let us consider the relations with three $\beta$s. As in the level-2 case, we have
\bea
&&\mathcal{V}(\beta_1,\beta_2,\beta_3;\alpha_1,...,\alpha_r|N)-\mathcal{V}(\beta_2,\beta_3;\beta_1,\alpha_1,...,\alpha_r|N)\nn
&=&\Sl_{\sigma\in P(O\{\alpha_1,...,\alpha_r\}\bigcup O\{\beta_1,\beta_2,\beta_3\})}\mathcal{P}_{\{\beta_3,\beta_2,\beta_1,\alpha,N\},\{\sigma,N\}}A_o(\sigma,N)\nn
&=&e^{-2i\pi\alpha'\left(k_{\beta_1}\cdot k_{\beta_2}+k_{\beta_2}\cdot k_{\beta_3}+k_{\beta_3}\cdot k_{\beta_1}\right)}\Sl_{\sigma\in P(O\{\alpha_1,...,\alpha_r\}\bigcup O\{\beta_1,\beta_2,\beta_3\})}\mathcal{P}_{\{\beta_1,\beta_2,\beta_3,\alpha,N\},\{\sigma,N\}}A_o(\sigma,N).
\eea
From this, we can express $\mathcal{U}(\beta_1,\beta_2,\beta_3,\alpha_1,...,\alpha_r|N)$ by $\mathcal{V}(\beta_1,\beta_2,\beta_3;\alpha_1,...,\alpha_r|N)$ and $\mathcal{V}(\beta_2,\beta_3;\beta_1,\alpha_1,...,\alpha_r|N)$ as
\bea
&&\mathcal{U}(\beta_1,\beta_2,\beta_3,\alpha_1,...,\alpha_r|N)\nn
&=&e^{2i\pi\alpha'\left(k_{\beta_1}\cdot k_{\beta_2}+k_{\beta_2}\cdot k_{\beta_3}+k_{\beta_3}\cdot k_{\beta_1}\right)}
\left[\mathcal{V}(\beta_1,\beta_2,\beta_3;\alpha_1,...,\alpha_r|N)-\mathcal{V}(\beta_2,\beta_3;\beta_1,\alpha_1,...,\alpha_r|N)\right].
\eea
We can define
\bea
&&a_3=e^{-2i\pi\alpha'\left(k_{\beta_1}\cdot k_{\beta_2}+k_{\beta_2}\cdot k_{\beta_3}+k_{\beta_3}\cdot k_{\beta_1}\right)}\mathcal{U}(\beta_1,\beta_2,\beta_3,\alpha_1,...,\alpha_r|N),\nn
&&b_3=\mathcal{V}(\beta_1,\beta_2,\beta_3;\alpha_1,...,\alpha_r|N), b_2=\mathcal{V}(\beta_2,\beta_3;\beta_1,\alpha_1,...,\alpha_r|N).
\eea
As we have shown in the cases of level-1 and level-2, we can define
\bea
&&a_2=-e^{-2i\pi\alpha'k_{\beta_2}\cdot k_{\beta_3}}\mathcal{U}(\beta_2,\beta_3;\beta_1,\alpha_1,...,\alpha_r|N),b_1=\mathcal{V}(\beta_3;\beta_2,\beta_1,\alpha_1,...,\alpha_r|N),\nn
&&a_1=\mathcal{U}(\beta_3;\beta_2,\beta_1,\alpha_1,...,\alpha_r|N),b_0=0.
\eea
Again, we have \eqref{recursion_1} and then from \eqref{recursion_2} we can express
$\mathcal{V}(\beta_1,\beta_2,\beta_3;\alpha_1,...,\alpha_r|N)$  by the $\mathcal{U}$s with not more than three$\beta$s
\bea
&&\mathcal{V}(\beta_1,\beta_2,\beta_3;\alpha_1,...,\alpha_r|N)\nn
&=&\mathcal{U}(\beta_3;\beta_2,\beta_1,\alpha_1,...,\alpha_r|N)-e^{-2i\pi\alpha'k_{\beta_2}\cdot k_{\beta_3}}\mathcal{U}(\beta_2,\beta_3;\beta_1,\alpha_1,...,\alpha_r|N)\nn
&&+e^{-2i\pi\alpha'\left(k_{\beta_1}\cdot k_{\beta_2}+k_{\beta_2}\cdot k_{\beta_3}+k_{\beta_3}\cdot k_{\beta_1}\right)}\mathcal{U}(\beta_1,\beta_2,\beta_3;\alpha_1,...,\alpha_r|N).
\eea
Therefore, the KK-BCJ relations with not more than three $\beta$s are equivalent with the generalized
$U(1)$-like decoupling identities with not more than three $\beta
$s. The KK-BCJ relation at level-3 thus can be generated by the primary relations.

\subsubsection{General discussion}

As what we have shown in the above examples, in general,
 from the definitions \eqref{lhs_gen_u1}, \eqref{lhs_KK_BCJ} and the property of momentum kernel \eqref{kernel_3},
 we have
\bea\label{connect_I_J}
&&\mathcal{V}(\beta_1,...,\beta_s;\alpha_1,...,\alpha_r|N)-\mathcal{V}(\beta_2,...,\beta_s;\beta_1,\alpha_1,...,\alpha_r|N)\nn
&=&(-1)^{s-1}e^{-2i\pi\alpha'\Sl_{1\leq i<j\leq s}k_{\beta_i}\cdot k_{\beta_j}}\mathcal{U}(\beta_1,...,\beta_s;\alpha_1,...,\alpha_r|N).
\eea
This formula express any $\mathcal{U}(\beta;\alpha|N)$ by $\mathcal{V}(\beta;\alpha|N)$s at level-$s$ and level-$(s-1)$.
To express $\mathcal{V}(\beta;\alpha|N)$ by $\mathcal{U}(\beta;\alpha|N)$s,
according to the above equation with no more than $s$ $\beta$s  in $\mathcal{U}(\beta;\alpha|N)$,
we can define
\bea
&&a_s=(-1)^{s-1}e^{-2i\pi\alpha'\Sl_{1\leq i<j\leq s}k_{\beta_i}\cdot k_{\beta_j}}
\mathcal{U}(\beta_1,...,\beta_s;\alpha_1,...,\alpha_r|N),\nn
&&b_s=\mathcal{V}(\beta_1,...,\beta_s;\alpha_1,...,\alpha_r|N),\nn
&&a_{s-t}=(-1)^{s-t-1}e^{-2i\pi\alpha'\Sl_{t+1\leq i<j\leq s}k_{\beta_i}\cdot k_{\beta_j}}\mathcal{U}(\beta_{t+1},...,\beta_s;\beta_t,...,\beta_1,\alpha_1,...,\alpha_r|N)(\text{for  } 1\leq t<s),\nn
&&b_{s-t}=\mathcal{V}(\beta_{t+1},...,\beta_s;\beta_t,...,\beta_1\alpha_1,...,\alpha_r|N) (\text{for  } 1\leq t<s),\nn
&& b_0=0.
\eea
Under this definition of $a$ and $b$, the relation \eqref{recursion_1} is satisfied. Thus according to \eqref{recursion_2}, we have
\bea
&&\mathcal{V}(\beta_1,...,\beta_s;\alpha_1,...,\alpha_r|N)\nn
&=&(-1)^{s-1}e^{-2i\pi\alpha'\Sl_{1\leq i<j\leq s}k_{\beta_i}\cdot k_{\beta_j}}\mathcal{U}(\beta_1,...,\beta_s;\alpha_1,...,\alpha_r|N)\nn
&+&\Sl_{t=1}^{s-1}(-1)^{s-t-1}e^{-2i\pi\alpha'\Sl_{t+1\leq i<j\leq s}k_{\beta_i}\cdot k_{\beta_j}}\mathcal{U}(\beta_{t+1},...,\beta_s;\beta_t,...,\beta_1,\alpha_1,...,\alpha_r|N).
\eea
Since the generalized $U(1)$-like decoupling identities have been generated by the primary relations,
the KK-BCJ relations $\mathcal{V}(\beta_1,...,\beta_s;\alpha_1,...,\alpha_r|N)=0$ also holds automatically.
This tells us all the KK-BCJ relations are generated by the primary relations.

Before we continue this discussions to the field theory limits, we should pay attention to the boundary case.
The boundary case of KK-BCJ relations with only one $\alpha$ is given as
\bea
\mathcal{V}(\beta_1,...,\beta_s;\alpha_1|N)=0.
\eea
This is nothing but just the color-order reversed relation
\bea
A_o(\beta_s,...,\beta_1,\alpha_1,N)=(-1)^{s}e^{-2i\pi\alpha'\left(\Sl_{1\leq i<j\leq s}k_{\beta_i}\cdot k_{\beta_j}+\Sl_{i=1}^s k_{\beta_i}\cdot k_{\alpha_1}\right)}A_o(\alpha_1,\beta_1,...,\beta_s,N).
\eea
Using momentum conservation, we have
\bea
&&\Sl_{1\leq i<j\leq s}2k_{\beta_i}\cdot k_{\beta_j}+\Sl_{i=1}^s k_{\beta_i}\nn
&=&(k_{\beta_1}+...+k_{\beta_s}+k_{\alpha_1})^2-k_{\beta_1}^2-...-k_{\beta_s}^2-k_{\alpha_1}^2\nn
&=&k_N^2-k_{\beta_1}^2-...-k_{\beta_s}^2-k_{\alpha_1}^2.
\eea
Using on-shell conditions and $m_i^2\in \mathbb{Z}$ for $i=1,...,N$, we can give the color-order reversed relation as
\footnote{This color-order reversed relations is a little different from the ordinary one
 $A_o(N,\alpha_1,\beta_s,...,\beta_1)=(-1)^{N}\prod_{i=1}^N(-1)^{\alpha'm_i^2}A_o(\beta_1,...,\beta_s,\alpha_1,N)$.
  In fact, cyclic symmetry connects this two expressions.}
\bea
A_o(\beta_s,...,\beta_1,\alpha_1,N)=(-1)^{N}\prod_{i=1}^N(-1)^{\alpha'm_i^2}A_o(\alpha_1,\beta_1,...,\beta_s,N).
\eea

\subsection{Field theory limits}

We can extend all above discussions in string theory to field theory by taking $\alpha'\rightarrow 0$.
 After taking the field theory limits, the massless states of open strings are left.
We get the relations among pure-gluon amplitudes.
Only the leading orders of the real part and the imaginary part contribute to the relations.
To write down the field theory limits of the monodromy relations, we should use the momentum kernel
\bea\label{kernel_f}
\mathcal{P}^f_{\{\sigma\},\{\tau\}}=1-i\alpha'2\pi\Sl_{i,j}k_i\cdot k_j\theta(\sigma^{-1}(i)-\sigma^{-1}(j))\theta(\tau^{-1}(j)-\tau^{-1}(i)).
\eea
To generate the KK and BCJ relations in field theory by primary relations, we should first
express the corresponding KK and BCJ relations in string theory by the primary relations.
 Then taking field theory limits $\alpha'\rightarrow 0$ in the relations. The difference
 from string theory is that the $U(1)$ decoupling identity(fundamental KK relation) cannot
 be chosen as one of the primary relations in field theory. This is because we need the kinematic
 factors $s_{ij}$ in BCJ relations in field theory, but in  KK relations in string theory,
 the kinematic factors
 only come  from higher-order terms of the expansion of cosine functions.
 When taking the field theory limits, we only keep the leading parts of the cosine functions,
 the kinematic factors are ignored. However,
Since there are kinematic factors in the BCJ relations in field theory,
we can consider the fundamental BCJ relation and the cyclic symmetry as the primary relations.

Now we consider a five-point example  with two $\beta$s to show the details.
 The five-point KK relation with two $\beta$s  is given as
\bea\label{eg_KK}
A_g(4,3,1,2,5)-A_g(1,3,4,2,5)-A_g(1,3,2,4,5)-A_g(1,2,3,4,5)=0,
\eea
where we consider the legs $3$ and $4$ as the $\beta_1$, $\beta_2$. Correspondingly,
the BCJ relation with two $\beta$s is given as
\bea\label{eg_BCJ}
&&(s_{31}+s_{41}+s_{43})A_g(1,3,4,2,5)+(s_{31}+s_{41}+s_{43}+s_{42})A_g(1,3,2,4,5)\nn
&+&(s_{31}+s_{41}+s_{43}+s_{42}+s_{32})A_g(1,2,3,4,5)=0.
\eea
We denote the L. H. S. of the  KK and BCJ relations in string theory as $\mathcal{K}(\beta;\alpha|N)$ and
$\mathcal{B}(\beta;\alpha|N)$ respectively. The field theory limits of the L.H.S. of KK and BCJ relations
are denoted by $\mathcal{K}^f(\beta;\alpha|N)$ and
$\mathcal{B}^f(\beta;\alpha|N)$.
To show the two relations \eqref{eg_KK} and \eqref{eg_BCJ} can be generated by the fundamental BCJ relation and the
cyclic symmetry, we can write this two relations into a complex KK-BCJ relation
\bea\label{5pt_complex}
&&\mathcal{K}^f(3,4;1,2|5)+i\pi\alpha'\mathcal{B}^f(3,4;1,2|5)\nn
&=&A_g(4,3,1,2,5)-[1-i\pi\alpha'(s_{31}+s_{41}+s_{43})]A_g(1,3,4,2,5)\nn
&-&[1-i\pi\alpha'(s_{31}+s_{41}+s_{43}+s_{42})]A_g(1,3,2,4,5)\nn
&-&[1-i\pi\alpha'(s_{31}+s_{41}+s_{43}+s_{42}+s_{32})]A_g(1,2,3,4,5)\nn
&=&0.
\eea
To show how to generate this relation by the primary relations,
we first express $\mathcal{V}(3,4;1,2|5)$ in string theory.
Using \eqref{level_2_V}, \eqref{level_2_U}, \eqref{2_by_1_U1}
and \eqref{2_by_1_U2}, we have
\bea\label{eg_complex}
&&\mathcal{V}(3,4;1,2|5)\nn
&=&\mathcal{U}(4;3,1,2|5)-e^{-i\pi\alpha's_{34}}\mathcal{U}(3,4;1,2|5)\nn
&=&\mathcal{U}(4;3,1,2|5)-e^{-i\pi\alpha's_{34}}\frac{1}{2i\sin(\pi\alpha's_{34})}
\left[e^{i\pi\alpha's_{34}}\mathcal{U}(3;4;1,2|5)-\mathcal{U}(4;3;1,2|5)\right]\nn
&=&\mathcal{U}(4;3,1,2|5)-\frac{1}{2i\sin(\pi\alpha's_{34})}\nn
&&\times\Bigl\{\mathcal{U}(3;4,1,2|5)+e^{-i\pi\alpha's_{41}}\mathcal{U}(3;1,4,2|5)
+e^{-i\pi\alpha'(s_{41}+s_{42})}\mathcal{U}(3;1,2,4|5)\nn
&&-e^{-i\pi\alpha's_{34}}\mathcal{U}(4;3,1,2|5)-e^{-i\pi\alpha'(s_{34}+s_{31})}\mathcal{U}(4;1,3,2|5)
-e^{-i\pi\alpha'(s_{34}+s_{31}+s_{32})}\mathcal{U}(4;1,2,3|5)\Bigr\}.
\eea
As we have pointed in the subsection \eqref{subsection_BCJ_cyclic},
when we consider the fundamental BCJ relation and the cyclic symmetry as the primary relations, we  have
\bea\label{fun_KK_in_fun_BCJ}
\mathcal{K}(\beta_1;\alpha_1,...,\alpha_r|N)=-\frac{1}{\sin(\pi\alpha's_{\beta_1\alpha_1})}\mathcal{B}(\beta_1;\alpha_2,...,\alpha_r,N|\alpha_1)
+\cot(\pi\alpha's_{\beta_1\alpha_1})\mathcal{B}(\beta_1;\alpha_1,...,\alpha_r|N),
\eea
where$\mathcal{K}(\beta_1;\alpha_1,...,\alpha_r|N)$ and  $\mathcal{B}(\beta_1;\alpha_1,...,\alpha_r|N)$
are the L. H. S. of the fundamental KK and the fundamental BCJ relations respectively.
Thus we can express $\mathcal{V}(3,4;1,2|5)$ by $\mathcal{B}$s with only one $\beta$ as
\bea\label{eg_primary}
&&\mathcal{V}(3,4;1,2|5)\nn
&=&\mathcal{K}(3,4;1,2|5)+i\mathcal{B}(3,4;1,2|5)\nn
&=&-\frac{1}{\sin(\pi\alpha's_{43})}\mathcal{B}(4;1,2,5|3)
+\cot(\pi\alpha's_{43})\mathcal{B}(4;3,1,2|5)-i\mathcal{B}(4;3,1,2|5)\nn
&&-\frac{1}{2i\sin(\pi\alpha's_{34})}
\times\Biggl\{\Biggl[-\frac{1}{\sin(\pi\alpha's_{34})}\mathcal{B}(3;1,2,5|4)
+\cot(\pi\alpha's_{34})\mathcal{B}(3;4,1,2|5)-i\mathcal{B}(3;4,1,2|5)\nn
&&+e^{-i\pi\alpha's_{41}}\left(-\frac{1}{\sin(\pi\alpha's_{31})}\mathcal{B}(3;4,2,5|1)
+\cot(\pi\alpha's_{31})\mathcal{B}(3;1,4,2|5)-i\mathcal{B}(3;1,4,2|5)\right)\nn
&&+e^{-i\pi\alpha'(s_{41}+s_{42})}\left(-\frac{1}{\sin(\pi\alpha's_{31})}\mathcal{B}(3;2,4,5|1)
+\cot(\pi\alpha's_{31})\mathcal{B}(3;1,2,4|5)-i\mathcal{B}(3;1,2,4|5)\right)\Biggr]\nn
&&-e^{-i\pi\alpha's_{34}}\left(-\frac{1}{\sin(\pi\alpha's_{43})}\mathcal{B}(4;1,2,5|3)
+\cot(\pi\alpha's_{43})\mathcal{B}(4;3,1,2|5)-i\mathcal{B}(4;3,1,2|5)\right)\nn
&&-e^{-i\pi\alpha'(s_{34}+s_{31})}\left(-\frac{1}{\sin(\pi\alpha's_{41})}\mathcal{B}(4;3,2,5|1)
+\cot(\pi\alpha's_{41})\mathcal{B}(4;1,3,2|5)-i\mathcal{B}(4;1,3,2|5)\right)\nn
&&-e^{-i\pi\alpha'(s_{34}+s_{31}+s_{32})}\left(-\frac{1}{\sin(\pi\alpha's_{41})}\mathcal{B}(4;2,3,5|1)
+\cot(\pi\alpha's_{41})\mathcal{B}(4;1,2,3|5)-i\mathcal{B}(4;1,2,3|5)\right)\Biggr\}.\nn
\eea
After taking $\alpha'\rightarrow0$, we should use $\mathcal{K}^f$ and $\pi\alpha'\mathcal{B}^f$,
instead of $\mathcal{K}$ and $\mathcal{B}$ respectively.

Expanding both sides of the above equation
according to the powers of $\alpha'$. We can write down the contributions from different
orders in $\alpha'$.
The first contribution of the above equation may be $(\alpha')^{-1}$. In this case, we have
\bea
&-&\frac{1}{2i\pi\alpha's_{34}}\Biggl[-\frac{1}{s_{34}}\mathcal{B}^f(3;1,2,5|4)
+\frac{1}{s_{34}}\mathcal{B}^f(3;4,1,2|5)-\frac{1}{s_{31}}\mathcal{B}^f(3;4,2,5|1)
+\frac{1}{s_{31}}\mathcal{B}^f(3;1,4,2|5)\nn
&&-\frac{1}{s_{31}}\mathcal{B}^f(3;2,4,5|1)+\frac{1}{s_{31}}\mathcal{B}^f(3;1,2,4|5)\nn
&&+\frac{1}{s_{43}}\mathcal{B}^f(4;1,2,5|3)-\frac{1}{s_{43}}\mathcal{B}^f(4;3,1,2|5)
+\frac{1}{s_{41}}\mathcal{B}^f(4;3,2,5|1)-\frac{1}{s_{41}}\mathcal{B}^f(4;1,3,2|5)\nn
&&+\frac{1}{s_{41}}\mathcal{B}^f(4;2,3,5|1)-\frac{1}{s_{41}}\mathcal{B}^f(4;1,2,3|5)\Biggr]
\eea
However, using the leading order of \eqref{fun_KK_in_fun_BCJ}, the above equation becomes
\bea
&&-\frac{1}{2i\pi\alpha's_{34}}\Bigl[\mathcal{K}^f(3;4,1,2|5)+\mathcal{K}^f(3;1,4,2|5)+\mathcal{K}^f(3;1,2,4|5)\nn
&&-\mathcal{K}^f(4;3,1,2|5)-\mathcal{K}^f(4;1,3,2|5)-\mathcal{K}^f(4;1,2,3|5)\Bigr].
\eea
This contribution vanishes because the first three terms can be obtained by first inserting
$4$ then inserting $3$ at the possible locations while the last three terms can be obtained
by first inserting $3$ then inserting $4$ at the possible the locations. The two kinds of
insertion are equivalent. Noticing the first three terms  and the last three terms have
different signs, they cancel out with each other.

Now we consider the $(\alpha')^0$ order. In the L. H. S. of \eqref{eg_primary}, it is just
the leading order of the real part of $\mathcal{K}(3,4;1,2|5)$, i. e., the L. H. S. of field
theory KK \eqref{eg_KK}. The R. H. S. of  \eqref{eg_primary} then expresses $\mathcal{K}^f(3,4;1,2|5)$
by the fundamental BCJ relations in field theory as
\bea
&&\mathcal{K}^f(3,4;1,2|5)\nn
&=&-\frac{1}{2s_{43}}\mathcal{B}^f(4;1,2,5|3)+\frac{1}{2s_{34}}\mathcal{B}^f(3;4,1,2|5)\nn
&&+\frac{1}{2s_{34}}\frac{s_{31}+s_{41}}{s_{31}}\mathcal{B}^f(3;1,4,2|5)
+\frac{1}{2s_{34}}\frac{s_{31}+s_{41}+s_{42}}{s_{31}}\mathcal{B}^f(3;1,2,4|5)\nn
&&-\frac{1}{2s_{34}}\frac{s_{41}+s_{34}+s_{31}}{s_{41}}\mathcal{B}^f(4;1,3,2|5)
-\frac{1}{2s_{34}}\frac{s_{41}+s_{34}+s_{31}+s_{32}}{s_{41}}\mathcal{B}^f(4;1,2,3|5)\nn
&&-\frac{1}{2s_{34}}\frac{s_{41}}{s_{31}}\mathcal{B}^f(3;4,2,5|1)
-\frac{1}{2s_{34}}\frac{1}{2s_{34}}\frac{s_{41}+s_{42}}{s_{31}}\mathcal{B}^f(3;2,4,5|1)\nn
&&+\frac{1}{2s_{34}}\frac{s_{34}+s_{31}}{s_{41}}\mathcal{B}^f(4;3,2,5|1)
+\frac{1}{2s_{34}}\frac{s_{34}+s_{31}+s_{32}}{s_{41}}\mathcal{B}^f(4;2,3,5|1).
\eea
One can check this equation by writing the $\mathcal{K}^f$ and the $\mathcal{B}^f$s explicitly in
 terms of amplitudes.
From this equation, we can see, the KK relation with two $\beta$s
is nothing but just a linear combination of the fundamental BCJ relations.

For the $\alpha'$ order, $\mathcal{V}(3,4;1,2|5)$ only has a term
$i\pi\alpha'\mathcal{B}^f(3,4;1,2|5)$ which is the leading order of the
 imaginary part of $\mathcal{V}(3,4;1,2|5)$.
Thus from \eqref{eg_primary}, we express the $\mathcal{B}^f(3,4;1,2|5)$ as linear combination of L. H. S. of fundamental BCJ relations by collecting the $\alpha'$ terms in
R. H. S. of \eqref{eg_primary}\footnote{In the derivation, we encounter the contributions from the
phase factors of the form $e^{i\alpha'A}$, the coefficients of KK relation and the coefficients of
BCJ relation. Since there is the factor $\frac{1}{2i\pi\alpha's_{34}}$, to get contribution of the
terms in the braces in\eqref{eg_complex}, we should keep the terms in $(\alpha')^2$ in the braces.
The terms with $(\alpha')^2$ may come from the expansion of phase factor with $(\alpha')^2$, the
expansion of the coefficients of KK relation with $(\alpha')^2$ and the $\alpha'$ term in the
expansion of phase factor multiplied by the $\alpha'$ term in BCJ relation. Since for each
$e^{i\alpha'A}(\cos(\alpha'B)+i\sin(\alpha'B))$ there is a corresponding term of the form
$-e^{i\alpha'(A+a)}(\cos(\alpha'(B+a))+i\sin(\alpha'(B+a)))$. Here $e^{i\alpha'A}$ is the phase
factor, cosine functions are the coefficients in KK relation, sine functions are the coefficients
in the BCJ relation, $a=s_{34}$. We can use $\frac{1}{2}A^2+\frac{1}{2}B^2-\frac{1}{2}(A+a)^2-
\frac{1}{2}(B+a)^2=AB-(A+a)(B+a)$ to write all the tree kinds of contributions by
BCJ coefficients. At last, we express $\mathcal{B}^f(3,4;1,2|5)$ by only fundamental BCJ relations.}
\bea
&&\mathcal{B}^f(3,4;1,2|5)\nn
&=&-\frac{s_{41}}{s_{34}}\mathcal{B}^f(3;1,4,2|5)-\frac{s_{41}+s_{42}}{s_{34}}\mathcal{B}^f(3;1,2,4|5)\nn
& &+\frac{s_{34}+s_{31}}{s_{34}}\mathcal{B}^f(4;1,3,2|5)+\frac{s_{34}+s_{31}+s_{32}}{s_{34}}\mathcal{B}^f(4;1,2,3|5).
\eea
This can also be checked by writing the $\mathcal{B}^f$s  explicitly in terms of amplitudes.
This equation expresses the BCJ relation with two $\beta$s for five-point amplitudes as a
linear combination of fundamental BCJ relations. Once we have the fundamental BCJ relation,
the BCJ relation with two $\beta$s holds naturally.


\section{General monodromy relation}\label{section_4}
In the previous section, we have seen the equivalence between
the KK-BCJ relations and generalized $U(1)$-like decoupling identities.
In this section, we extend this discussion to a more general case. We
will give the general monodromy relations\eqref{generalized_KK_BCJ}.
To show \eqref{generalized_KK_BCJ}, we define a linear combination of
amplitudes $\mathcal{W}(\gamma;\beta;\alpha|N)$ as the L. H. S. of \eqref{generalized_KK_BCJ}
\bea
& &\mathcal{W}(\gamma_1,...,\gamma_t;\beta_1,...,\beta_s;\alpha_1,...,\alpha_r|N)\nn
&\equiv&\Sl_{\tau\in P(O\{\gamma\}\bigcup O\{\beta^T\})}\mathcal{P}^*_{\{\gamma,\beta^T,\alpha,N\},\{\tau,\alpha,N\}}A_o(\tau,\alpha,N)\nn
&+&(-1)^{s-1}\Sl_{\sigma\in P(O\{\alpha\}\bigcup O\{\beta\})|\sigma_1=\alpha_1}\mathcal{P}_{\{\gamma,\beta^T,\alpha,N\},\{\gamma,\sigma,N\}}A_o(\gamma,\sigma,N).\nn
\eea
 \eqref{generalized_KK_BCJ} in this definition becomes
 $\mathcal{W}(\gamma_1,...,\gamma_t;\beta_1,...,\beta_s;\alpha_1,...,\alpha_r|N)=0$.
This kind of relation can be seen from monodromy and can also be generated by the primary relations.
 After taking the field theory limits, we can obtain the corresponding relations in field theory.

 \subsection{Monodromy}

Now we will show that \eqref{generalized_KK_BCJ} is the result of monodromy.
To show this, we first return to the monodromy derivation of KK-BCJ relations  \eqref{KK_BCJ}
\cite{BjerrumBohr:2009rd, Stieberger:2009hq, BjerrumBohr:2010zs}. We consider
an open string worldsheet integral. In this integral  the vertex operators of the
legs $\beta_i$ has the relative ordering $0<Z_{\beta_1}<Z_{\beta_2}<...<Z_{\beta_s}$
for $Z_{\beta_s}>0$ and $Z_{\beta_s}<Z_{\beta_{s-1}}<...<Z_{\beta_1}<0$ for
$Z_{\beta_s}<0$, while the vertex operators of legs $\alpha_j$s has  the relative ordering
$Z_{\alpha_1}<Z_{\alpha_2}<...<Z_{\alpha_r}$. We have the integral
\bea\label{KK_BCJ_integral}
&&\langle \phi_N|\int\limits_{1}^{\infty}d Z_{\alpha_{r}}V_{\alpha_r}(Z_{\alpha_r})...\int\limits_{{1}}^{Z_{\alpha_4}}dZ_{\alpha_{3}}V_{\alpha_3}(Z_{\alpha_3})
V_{\alpha_2}(Z_{\alpha_2=1})\nn
&&\cdot
\int\limits_{-\infty+i\epsilon}^{\infty+i\epsilon}dZ_{\beta_s} V_{\beta_s}(Z_{\beta_s})\int\limits_{0}^{1}d\rho_{\beta_{s-1}}Z_{\beta_s} V_{\beta_s}(\rho_{\beta_{s-1}}Z_{\beta_s})...\int\limits_{0}^{\rho_{\beta_2}}d\rho_{\beta_1}Z_{\beta_s} V_{\beta_1}(\rho_{\beta_1}Z_{\beta_s})|\phi_{\alpha_1}\rangle\nn
&=&\langle \phi_N|\int\limits_{1}^{\infty}d Z_{\alpha_{r}}V_{\alpha_r}(Z_{\alpha_r})...\int\limits_{{1}}^{Z_{\alpha_4}}dZ_{\alpha_{3}}V_{\alpha_3}(Z_{\alpha_3})
V_{\alpha_2}(Z_{\alpha_2}=1)
\int\limits_{-\infty+i\epsilon}^{\infty+i\epsilon}dZ_{\beta_s}\mathcal{O}_{\beta}(Z_{\beta_s},0)|\phi_{\alpha_1}\rangle,
\eea
where $\rho_{\beta_i}$ for a given $i~(1\leq i<s)$ is defined as $\rho_{\beta_i}=\frac{Z_{\beta_i}}{Z_{\beta_s}}$,
 thus the $Z_{\beta_i}$ integrals are written as $\rho_{\beta_i}$ integrals.
In the above equation, we have fixed $Z_{\alpha_2}=1$, the two states $\langle\phi_N|$ and $|\phi_{\alpha_1}\rangle$ are inserted at $Z_N=\infty$ and $Z_{\alpha_1}=0$ respectively. The operator $\mathcal{O}$ is defined as
\bea
\mathcal{O}_{\beta}(Z_{\beta_s},0)=V_{\beta_s}(Z_{\beta_s})\int\limits_{0}^{1}d\rho_{\beta_{s-1}}Z_{\beta_s} V_{\beta_s}(\rho_{\beta_{s-1}}Z_{\beta_s})...\int\limits_{0}^{\rho_{\beta_2}}d\rho_{\beta_1}Z_{\beta_s} V_{\beta_1}(\rho_{\beta_1}Z_{\beta_s}).
\eea
\begin{figure}
\begin{center}
\includegraphics[width=0.8\textwidth]{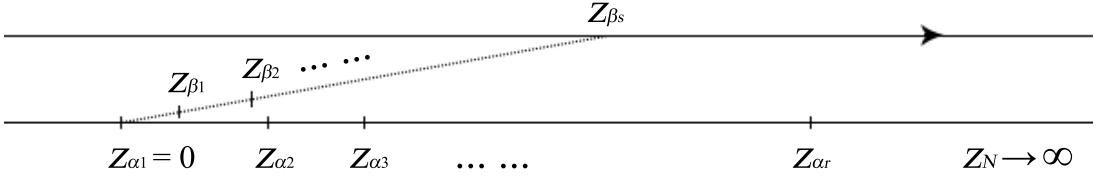}
\end{center}
\caption{Contour for KK-BCJ relation}
\label{fig:1}
\end{figure}
This $Z_{\beta_s}$ integral can be expressed by Fig. \ref{fig:1} and it must vanish
when we close the contour above the real axis. We consider  $Z_{\alpha_1}=0<Z_{\beta_s}<Z_{\alpha_2}=1$ as a
standard integral region, when we break the worldsheet integral into pieces corresponding to different
possible permutations of all legs, for a given integral ordering, we can adjust the positions of vertex
operators to make the them in the same ordering with the corresponding integrals. Then we get
the amplitude with legs in this ordering. Considering the branch point are at the positions of vertex operators,
when we move an vertex operator $V_{\beta_i}$ from the left side to the right side of another vertex operator
$V_j$, comparing to the former ordering, the amplitude with legs in the latter ordering should be accompanied by a phase
factor $e^{2\alpha'i\pi k_{\beta_i}\cdot k_j}$. Similarly, if we move $V_{\beta_i}$ from the right side
to the left side of another vertex operator $V_j$, we obtain a phase factor
$e^{-2\alpha'i\pi k_{\beta_i}\cdot k_j}$. After considering all the phase factors, The vanishing of the
integral \eqref{KK_BCJ_integral} gives \eqref{KK_BCJ}. Particularly, the integral
with $Z_{\beta_s}<Z_{\beta_{s-1}}<...<Z_{\beta_1}<0$ corresponds to the first term of \eqref{KK_BCJ}
while the integrals with  $0<Z_{\beta_1}<Z_{\beta_2}<...<Z_{\beta_s}$ correspond to the second term
of \eqref{KK_BCJ}.

\begin{figure}
\begin{center}
\includegraphics[width=0.8\textwidth]{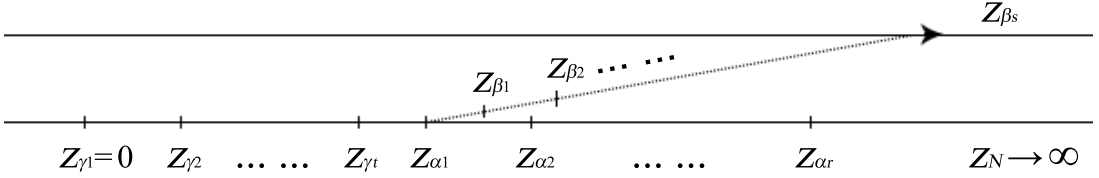}
\end{center}
\caption{Contour for general monodromy relation}
\label{fig:2}
\end{figure}

Now we consider another vanishing worldsheet integral
\bea\label{general_monodromy_integral}
0&=&\langle\phi_N|\int\limits_{1}^{\infty}d Z_{\alpha_{r}}V_{\alpha_r}(Z_{\alpha_r})...\int\limits_{{Z_{\alpha_1}=1}}^{Z_{\alpha_{3}}}dZ_{\alpha_{2}}V_{\alpha_{2}}(Z_{\alpha_{2}})
\int\limits_{-\infty+i\epsilon}^{\infty+i\epsilon}dZ_{\beta_s}\mathcal{O}_{\beta}(Z_{\beta_s},1)V_{\alpha_1}(Z_{\alpha_1}=1)\nn
&&\cdot
\int\limits_{0}^{Z_{\alpha_1}=1}d Z_{\gamma_{t}}V_{\gamma_{t}}(Z_{\gamma_{t}})...\int\limits_{0}^{Z_{\gamma_{3}}}dZ_{\gamma_{2}}V_{\gamma_{2}}(Z_{\gamma_{2}})|\phi_{\gamma_1}\rangle,
\eea
where we have choose the fixed point as $Z_{\gamma_1}=0$, $Z_{\alpha_1}=1$ and $Z_N=\infty$. The operator $\mathcal{O}_{\beta}(Z_{\beta_s},x)$ is the extension of $\mathcal{O}_{\beta}(Z_{\beta_s},0)$
\bea
&&\mathcal{O}_{\beta}(Z_{\beta_s},x)\nn
&=&V_{\beta_s}(Z_{\beta_s})\int\limits_{0}^{1}d\rho_{\beta_{s-1}}(Z_{\beta_s}-x) V_{\beta_s}(\rho_{\beta_{s-1}}(Z_{\beta_s}-x)+x)...\int\limits_{0}^{\rho_{\beta_2}}d\rho_{\beta_1}(Z_{\beta_s}-x) V_{\beta_1}(\rho_{\beta_1}(Z_{\beta_s}-x)+x).
\eea
Here $\rho_{\beta_i}=\frac{Z_{\beta_i}-x}{Z_{\beta_s}-x}$. This integral can be expressed by Fig. \ref{fig:2}.
We can break all the integrals \eqref{general_monodromy_integral} into pieces corresponding to all the possible
 orderings of the legs and adjust the positions of the vertex operators.
As pointed in the case of KK-BCJ relation, an appropriate phase factor should be considered when we adjust
the positions of the vertex operators. Then for the permutations with the relative ordering
$Z_{\beta_s}<Z_{\beta_{s-1}}<...<Z_{\beta_1}<Z_{\alpha_1}=1$, we get the first term of \eqref{generalized_KK_BCJ},
 while for the permutations with the relative ordering $Z_{\alpha_1}=1<Z_{\beta_1}<Z_{\beta_2}<...<Z_{\beta_s}$,
 we get the second term of \eqref{generalized_KK_BCJ}. Thus we get the general monodromy relations
 \eqref{generalized_KK_BCJ}. In next subsection, we will see, this general monodromy relations can also be constructed
  by the primary relations.

 \subsection{Generating general monodromy relation by primary relations}
In this subsection, we will show that the general monodromy relations \eqref{generalized_KK_BCJ}
can also be generated by the primary relations. To see this, we just use the KK-BCJ relations to
generate this kind of relation. Since the KK-BCJ relations can be generated by the primary relations,
\eqref{generalized_KK_BCJ} can also be generated by the primary relations.

\subsubsection{An example}

 The boundary case of \eqref{generalized_KK_BCJ}  is $s=0$. When $s=0$, there is no $\beta$,
from the definition of $\mathcal{W}$, we have
\bea
\mathcal{W}(\gamma_1,...,\gamma_t;\emptyset;\alpha_1,...,\alpha_r|N)=A_o(\gamma,\alpha,N)-A_o(\gamma,\alpha,N)=0.
\eea
Now let us consider the \eqref{generalized_KK_BCJ} with $s=1$.

{\bf Level-1}

If there is only one $\beta$, \eqref{generalized_KK_BCJ} becomes
\bea\label{general_monodromy_eg}
&&\mathcal{W}(\gamma_1,...,\gamma_t;\beta_1;\alpha_1,...,\alpha_r|N)\nn
&=&\Sl_{\tau\in P(O\{\gamma\}\bigcup O\{\beta_1\})}\mathcal{P}^*_{\{\gamma,\beta_1,\alpha,N\},\{\tau,\alpha,N\}}A_o(\tau,\alpha,N)\nn
&+&(-1)^0\Sl_{\sigma\in P(O\{\alpha\}\bigcup O\{\beta_1\})|\sigma_1=\alpha_1}\mathcal{P}_{\{\gamma,\beta_1,\alpha,N\},\{\gamma,\sigma,N\}}A_o(\gamma,\sigma,N).\nn
\eea
The first element in the permutation $\tau'$ in $\Sl_{\tau'\in P( O\{\gamma,\alpha_1\}\bigcup O\{\beta_1\})}
 \mathcal{P}^*_{\{\gamma,\alpha_1,\beta_1,\alpha_2,...,\alpha_r,N\},
 \{\tau',\alpha_2,...,\alpha_r,N\}}A_o(\tau',\alpha_2,...,\alpha_r,N)$ can be either $\alpha_1$ or $\beta_1$.
 Thus the first term of
  can be written as \eqref{general_monodromy_eg}
\bea
&&\Sl_{\tau\in P(O\{\gamma\}\bigcup O\{\beta_1\})}\mathcal{P}^*_{\{\gamma,\beta_1,\alpha,N\},\{\tau,\alpha,N\}}A_o(\tau,\alpha,N)\nn
&=&e^{-2i\pi\alpha'k_{\alpha_1}\cdot k_{\beta_1}}\Biggl[\Sl_{\tau'\in P( O\{\gamma,\alpha_1\}\bigcup O\{\beta_1\})}\mathcal{P}^*_{\{\gamma,\alpha_1,\beta_1,\alpha_2,...,\alpha_r,N\},\{\tau',\alpha_2,...,\alpha_r,N\}}A_o(\tau',\alpha_2,...,\alpha_r,N)\nn
&&-\Sl_{\tau''\in P( O\{\gamma,\alpha_1\})}\mathcal{P}^*_{\{\gamma,\alpha_1,\beta_1,\alpha_2,...,\alpha_r,N\},\{\tau'',\beta_1,\alpha_2,...,\alpha_r,N\}}
A_o(\tau'',\beta_1,\alpha_2,...,\alpha_r,N)\Biggr],
\eea
where \eqref{kernel_4} has been used.

Similarly, the second element in the permutation $\sigma$ in $\Sl_{\sigma\in P(O\{\alpha\}\bigcup O\{\beta_1\})|\sigma_1=\alpha_1}$
 can be either $\alpha_2$ or $\beta_1$. Thus
 we can write down the second term of \eqref{general_monodromy_eg} as
\bea
&&\Sl_{\sigma\in P(O\{\alpha\}\bigcup O\{\beta_1\})|\sigma_1=\alpha_1}\mathcal{P}_{\{\gamma,\beta_1,\alpha,N\},\{\gamma,\sigma,N\}}A_o(\gamma,\sigma,N)\nn
&=&e^{-2i\pi\alpha'k_{\alpha_1}\cdot k_{\beta_1}}\Biggl[\Sl_{\sigma'\in P(O\{\alpha_2,...,\alpha_r\}\bigcup O\{\beta_1\})|\sigma'_1=\alpha_2}\mathcal{P}_{\{\gamma,\alpha_1,\beta_1,\alpha_2,...,\alpha_r,N\},\{\gamma,\alpha_1,\sigma',N\}}A_o(\gamma,\alpha_1,\sigma',N)\nn
&+&\Sl_{\sigma''\in P(O\{\beta_1,\alpha_2,...,\alpha_r\})|\sigma''_1=\beta_1}\mathcal{P}_{\{\gamma,\alpha_1,\beta_1,\alpha_2,...,\alpha_r,N\},\{\gamma,\alpha_1,\sigma'',N\}}
A_o(\gamma,\alpha_1,\sigma'',N)\Biggr].
\eea
Considering both two terms, from the definition of $\mathcal{W}$, we get
\bea
&&\mathcal{W}(\gamma_1,...,\gamma_t;\beta_1;\alpha_1,...,\alpha_r|N)\nn
&=&e^{-2i\pi\alpha'k_{\alpha_1}\cdot k_{\beta_1}}\Biggl[\mathcal{W}(\gamma_1,...,\gamma_t,\alpha_1;\beta_1;\alpha_2,...,\alpha_r|N)
-\mathcal{W}(\gamma_1,...,\gamma_t,\alpha_1;\emptyset;\beta_1,\alpha_2,...,\alpha_r|N)\Biggr].\nn
\eea
Since the $\mathcal{W}$ in the boundary case with $s=0$ vanishes, we can express all
the $\mathcal{W}$s with $t$ $\gamma$s by those with $t-1$ $\gamma$s
\bea
\mathcal{W}(\gamma_1,...,\gamma_t;\beta_1;\alpha_2,...,\alpha_r|N)=e^{2i\pi\alpha'k_{\gamma_t}\cdot k_{\beta_1}}\mathcal{W}(\gamma_1,...,\gamma_{t-1};\beta_1;\gamma_t,\alpha_1,...,\alpha_r|N).
\eea
This gives a recursive relation. With this relation, we can express the
$\mathcal{W}$ with $t$ $\gamma$s by the $\mathcal{W}$ with no $\gamma$
\bea
\mathcal{W}(\gamma_1,...,\gamma_t;\beta_1;\alpha_2,...,\alpha_r|N)=e^{2i\pi\alpha'\Sl_{i=1}^tk_{\gamma_i}\cdot k_{\beta_1}}\mathcal{W}(\emptyset;\beta_1;\gamma_{1},...,\gamma_t,\alpha_1,...,\alpha_r|N).
\eea
Because we have $\mathcal{W}(\emptyset;\beta_1;\gamma_{1},...,\gamma_t,\alpha_1,...,\alpha_r|N)
=\mathcal{V}(\beta_1;\gamma_{1},...,\gamma_t,\alpha_1,...,\alpha_r|N)$,
using the KK-BCJ relation with one $\beta$(i.e., the $U(1)$-like decoupling identity),
we get the general monodromy relation with only one $\beta$
\bea
\mathcal{W}(\gamma_1,...,\gamma_t;\beta_1;\alpha_2,...,\alpha_r|N)=0.
\eea
\subsubsection{General proof}
Now we consider the general case with arbitrary number of $\beta$s.
As pointed in the above example, The first element in the permutation $\tau$ in $\Sl_{\tau'\in P(
O\{\gamma,\alpha_1\}\bigcup O\{\beta^T\})}\mathcal{P}^*_{\{\gamma,\alpha_1,\beta^T,\alpha_2,...,\alpha_r,N\},
\{\tau',\alpha_2,...,\alpha_r,N\}}A_o(\tau',\alpha_2,...,\alpha_r,N)$ can be either $\alpha_1$ or $\beta_1$.
Thus the first term of \eqref{generalized_KK_BCJ} can be expressed
 as
\bea
&&\Sl_{\tau\in P(O\{\gamma\}\bigcup O\{\beta^T\})}\mathcal{P}^*_{\{\gamma,\beta^T,\alpha,N\},\{\tau,\alpha,N\}}A_o(\tau,\alpha,N)\nn
&=&e^{-2i\pi\alpha'\Sl_{i=1}^sk_{\alpha_1}\cdot k_{\beta_i}}\Biggl[\Sl_{\tau'\in P( O\{\gamma,\alpha_1\}\bigcup O\{\beta^T\})}\mathcal{P}^*_{\{\gamma,\alpha_1,\beta^T,\alpha_2,...,\alpha_r,N\},\{\tau',\alpha_2,...,\alpha_r,N\}}A_o(\tau',\alpha_2,...,\alpha_r,N)\nn
&-&\Sl_{\tau''\in P( O\{\gamma,\alpha_1\}\bigcup O\{\beta_s,...,\beta_2\})}\mathcal{P}^*_{\{\gamma,\alpha_1,\beta_s,...,\beta_2,\beta_1\alpha_2,...,\alpha_r,N\},
\{\tau'',\beta_1,\alpha_2,...,\alpha_r,N\}}A_o(\tau'',\beta_1,\alpha_2,...,\alpha_r,N)\Biggr].
\eea
The second element in the permutation $\sigma$ in $\Sl_{\sigma\in P(O\{\alpha\}\bigcup O\{\beta\})|\sigma_1=\alpha_1}\mathcal{P}_{\{\gamma,\beta^T,\alpha,N\},\{\gamma,\sigma,N\}}A_o(\gamma,\sigma,N)$ can be
either $\alpha_2$ or $\beta_1$. Thus the second term of \eqref{generalized_KK_BCJ} becomes
\bea
&&\Sl_{\sigma\in P(O\{\alpha\}\bigcup O\{\beta\})|\sigma_1=\alpha_1}\mathcal{P}_{\{\gamma,\beta^T,\alpha,N\},\{\gamma,\sigma,N\}}A_o(\gamma,\sigma,N)\nn
&=&e^{-2i\pi\alpha'\Sl_{i=1}^sk_{\alpha_1}\cdot k_{\beta_i}}\Biggl[\Sl_{\sigma'\in P(O\{\alpha_2,...,\alpha_r\}\bigcup O\{\beta\})|\sigma'_1=\alpha_2}\mathcal{P}_{\{\gamma,\alpha_1,\beta^T,\alpha_2,...,\alpha_r,N\},\{\gamma,\alpha_1,\sigma',N\}}A_o(\gamma,\alpha_1,\sigma',N)\nn
&+&\Sl_{\sigma''\in P(O\{\beta_1,\alpha_2,...,\alpha_r\}\bigcup O\{\beta_2,...,\beta_s\})|\sigma''_1=\beta_1}\mathcal{P}_{\{\gamma,\alpha_1,\beta_s,...,\beta_2,\beta_1,\alpha_2,...,\alpha_r,N\},\{\gamma,\alpha_1,\sigma'',N\}}
A_o(\gamma,\alpha_1,\sigma'',N)\Biggr].
\eea
As what we have shown in the level-1 case, we have
\bea
&&\mathcal{W}(\gamma_1,...,\gamma_t;\beta_1,...,\beta_s;\alpha_1,...,\alpha_r|N)\nn
&=&e^{-2i\pi\alpha'\Sl_{i=1}^sk_{\alpha_1}\cdot k_{\beta_i}}\Biggl[\mathcal{W}(\gamma_1,...,\gamma_t,\alpha_1;\beta_1,...,\beta_s;\alpha_2,...,\alpha_r|N)
-\mathcal{W}(\gamma_1,...,\gamma_t,\alpha_1;\beta_2,...,\beta_s;\beta_1,\alpha_2,...,\alpha_r|N)\Biggr].\nn
\eea
With this relation, we can relate the $\mathcal{W}$s  which have $t-1$ $\gamma$s with those have $t$ $\gamma$s
\bea
&&e^{2i\pi\alpha'\Sl_{i=1}^sk_{\gamma_t}\cdot k_{\beta_i}}\mathcal{W}(\gamma_1,...,\gamma_{t-1};\beta_1,...,\beta_s;\gamma_t,\alpha_1,...,\alpha_r|N)\nn
&=&\Biggl[\mathcal{W}(\gamma_1,...,\gamma_t;\beta_1,...,\beta_s;\alpha_1,...,\alpha_r|N)
-\mathcal{W}(\gamma_1,...,\gamma_t;\beta_2,...,\beta_s;\beta_1,\alpha_1,...,\alpha_r|N)\Biggr].\nn
\eea
If we define
\bea
a_{s}&=&e^{2i\pi\alpha'\Sl_{i=1}^sk_{\gamma_t}\cdot k_{\beta_i}}
\mathcal{W}(\gamma_1,...,\gamma_{t-1};\beta_1,...,\beta_s;\gamma_t,\alpha_1,...,\alpha_r|N)\nn
b_s&=&\mathcal{W}(\gamma_1,...,\gamma_t;\beta_1,...,\beta_s;\alpha_1,...,\alpha_r|N)\nn
a_{s-j}&=&e^{2i\pi\alpha'\Sl_{i=j+1}^sk_{\gamma_t}\cdot k_{\beta_i}}\mathcal{W}(\gamma_1,...,\gamma_{t-1};\beta_{j+1},...,\beta_s;\gamma_t,\beta_j,...,\beta_1,\alpha_1,...,\alpha_r|N),(0< j<s)\nn
b_{s-j}&=&\mathcal{W}(\gamma_1,...,\gamma_t;\beta_{j+1},...,\beta_s;\beta_j,...,\beta_1,\alpha_1,...,\alpha_r|N),(0<j<s)\nn
b_0&=&\mathcal{W}(\gamma_1,...,\gamma_t;\emptyset;\beta_s,...,\beta_1,\alpha_1,...,\alpha_r|N).
\eea
The recursive relation \eqref{recursion_1} is satisfied again, thus from \eqref{recursion_2}
we can express the $\mathcal{W}$s with $t$ $\gamma$s by those with $t-1$ $\gamma$s
\bea
&&\mathcal{W}(\gamma_1,...,\gamma_t;\beta_1,...,\beta_s;\alpha_1,...,\alpha_r|N)\nn
&=&e^{2i\pi\alpha'\Sl_{i=1}^sk_{\gamma_t}\cdot k_{\beta_i}}\mathcal{W}(\gamma_1,...,\gamma_{t-1};\beta_1,...,\beta_s;\gamma_t,\alpha_1,...,\alpha_r|N)\nn
&+&\Sl_{j=1}^{s-1}e^{2i\pi\alpha'\Sl_{i=j+1}^sk_{\gamma_t}\cdot k_{\beta_i}}\mathcal{W}(\gamma_1,...,\gamma_{t-1};\beta_{j+1},...,\beta_s;\gamma_t,\beta_j,...,\beta_1,\alpha_1,...,\alpha_r|N).
\eea
The $\mathcal{W}$s with $t$ $\gamma$s and the $\mathcal{W}$s with $t-1$ $\gamma$s can be solved from
each other, thus they are equivalent relations. Using this recursive relation, we can express all the
$\mathcal{W}$s by linear combinations of those with no $\gamma$.
Since the boundary condition is given as
\bea
\mathcal{W}(\emptyset;\beta_1,...,\beta_s;\alpha_1,...,\alpha_r|N)=\mathcal{V}(\beta_1,...,\beta_s;\alpha_1,...,\alpha_r|N),
\eea
and the KK-BCJ relation $\mathcal{V}(\beta_1,...,\beta_s;\alpha_1,...,\alpha_r|N)=0$ can be generated by the
primary relations,
the general monodromy relations \eqref{generalized_KK_BCJ} thus can also be generated by the primary relations.
In the following subsection, we will discuss on the field theory limits of the
general monodromy relations \eqref{generalized_KK_BCJ}.

\subsection{Field theory limits}

We can obtain the field theory limits of the general monodromy relations \eqref{generalized_KK_BCJ}
by taking $\alpha'\rightarrow 0$ and replacing the definition of momentum kernel  by \eqref{kernel_f}.
Keeping the leading terms of both the real parts and the imaginary parts of the relations, we get the
corresponding relations
in field theory.

We just give an example here, when we consider the six-gluon amplitudes with $\{\gamma\}=\{1,2\}$, $\{\beta\}=\{3,4\}$, $\{\alpha\}=\{5\}$ and $N=6$. The $\mathcal{W}$ in field theory
is $\mathcal{W}^f$
\bea
&&\mathcal{W}^f(1,2;3,4;5|6)\nn
&=&A_g(1,2,4,3,5,6)\nn
&&+[1+i\pi\alpha'(s_{42}+s_{32})]A_g(1,4,3,2,5,6)
+[1+i\pi\alpha'(s_{42})]A_g(1,4,2,3,5,6)\nn
&&+[1+i\pi\alpha'(s_{41}+s_{31}+s_{42}+s_{32})]A_g(4,3,1,2,5,6)+[1+i\pi\alpha'(s_{41}+s_{42}+s_{32})]A_g(4,1,3,2,5,6)\nn
&&+[1+i\pi\alpha'(s_{41}+s_{42})]A_g(4,1,2,3,5,6)\nn
&&-[1-i\pi\alpha'(s_{45}+s_{34}+s_{35})]A_g(1,2,5,3,4,6)\nn
&&=0.
\eea
The real part of the above relation reads
\bea
&&A_g(1,2,4,3,5,6)+A_g(1,4,3,2,5,6)+A_g(1,4,2,3,5,6)\nn
&&+A_g(4,3,1,2,5,6)+A_g(4,1,3,2,5,6)+A_g(4,1,2,3,5,6)-A_g(1,2,5,4,3,6)\nn
&&=0.
\eea
This relation can be checked when we use KK relations to express
$A_g(4,3,1,2,5,6)$, $A_g(4,1,3,2,5,6)$ and $A_g(4,1,2,3,5,6)$
by the amplitudes with $1$ as the first leg and $6$ as the last leg.
The imaginary part relation is given as
\bea
&&(s_{42}+s_{32})A_g(1,4,3,2,5,6)+s_{42}A_g(1,4,2,3,5,6)+(s_{41}+s_{31}+s_{42}+s_{32})A_g(4,3,1,2,5,6)\nn
&&+(s_{41}+s_{42}+s_{32})A_g(4,1,3,2,5,6)
+(s_{41}+s_{42})A_g(4,1,2,3,5,6)+(s_{45}+s_{34}+s_{35})A_g(1,2,5,3,4,6)\nn
&&=0.
\eea
This one can also be checked by using BCJ relations and KK relations. In general, all other relations
corresponding to the real parts and the imaginary parts of the general monodromy relations can be given
in a similar way.
As in the field theory case of the KK and BCJ relations, both the real parts and the imaginary parts of
the field theory limits of general monodromy relations can be generated by the fundamental BCJ relations
and the cyclic symmetry.

\section{The minimal-basis expansion for color-ordered open string disk amplitudes}\label{section_5}

By the use of cyclic symmetry, KK and BCJ relations, one can reduce the number of independent amplitudes
from $N!$ to $(N-3)!$. The explicit expression of the minimal-basis expansion which expresses the KK basis
by the BCJ basis in field theory was conjectured in \cite{Bern:2008qj}.
It was pointed in \cite{Feng:2010my, Jia:2010nz} that the minimal-basis expansion could be solved
from a set of fundamental BCJ relations.
The conjectured formula \cite{Bern:2008qj} was proven in \cite{Chen:2011jx}. However,
the explicit minimal-basis expansion in string theory has not been given yet. In this section,
we will derive the minimal-basis expansion of color-ordered open string tree amplitudes. The field theory
limit of this expression gives the minimal-basis expansion
for color-ordered pure-gluon tree amplitudes.
In the following derivation, we  only use the BCJ relations which are the imaginary relation
of KK-BCJ relations \eqref{KK_BCJ}
\bea
\Sl_{\sigma\in P(O\{\beta\}\bigcup O\{2,\alpha\})}\mathcal{S}_{\{\beta_s,...,\beta_1,1,2,\alpha_1,...,\alpha_{N-s-3},N\},\{1,\sigma,N\}}
A_o(1,\sigma,N)=0,
\eea
where
\bea
\mathcal{S}_{\{\tau\},\{\sigma\}}=-\mathcal{I}m\mathcal{P}_{\{\tau\},\{\sigma\}}.
\eea
We will first show some examples.
\subsection{Examples}
In this subsection, we will show some examples.

{\bf Level-1}

If there is only one $\beta$, we return to the fundamental BCJ relation
\bea\label{min-basis-e-1}
&&A_o(1,\beta_1,2,\alpha_1,...,\alpha_{N-4},N)\nn
&=&-\frac{1}{\sin[\pi\alpha's_{1\beta_1}] }\Sl_{\sigma\in P(O\{\beta_1\}\bigcup O\{\alpha\})}
\mathcal{S}_{\{\beta_1,1,2,\alpha_1,...,\alpha_{N-4},N\},\{1,2,\sigma,N\}}A_o(1,2,\sigma,N).\nn
\eea
Here we have used the permutations in $P(O\{\beta_1\}\bigcup O\{\alpha\})$ are same with those in $P(\{\beta_1\}\bigcup O\{\alpha\})$.

{\bf Level-2}

The next example is given as the amplitudes with two $\beta$s.
If there are two $\beta$s, we have
\bea
&&A_o(1,\beta_1,\beta_2,2,\alpha_1,...,\alpha_{N-5},N)\nn
&=&-\frac{1}{\sin[\pi\alpha's_{1\beta_1\beta_2}]}\nn
&\times&\Biggl[\Sl_{\sigma\in P(O\{\beta_2\}\bigcup O\{\alpha\})}\mathcal{S}_{\{\beta_2,\beta_1,1,2,\alpha_1,...,\alpha_r,N\},\{1,\beta_1,2,\sigma,N\}}A_o(1,\beta_1,2,\sigma,N)\nn
&+&\Sl_{\tau\in P(O\{\beta_1,\beta_2\}\bigcup O\{\alpha\})}\mathcal{S}_{\{\beta_2,\beta_1,1,2,\alpha_1,...,\alpha_r,N\},\{1,2,\tau,N\}}A_o(1,2,\tau,N)\Biggr].
\eea
The amplitudes in the first term can be further expressed by the minimal-basis expansion with only one $\beta$.
Then we have
\bea\label{min-basis-e-2a}
&&A_o(1,\beta_1,\beta_2,2,\alpha_1,...,\alpha_{N-5},N)\nn
&=&-\frac{1}{\sin(\pi\alpha's_{1\beta_1\beta_2})}\Biggl[-\frac{1}{\sin(\pi\alpha's_{1\beta_1})}\nn
&&\times\Sl_{\sigma\in P(O\{\beta_2\}\bigcup O\{\alpha\})}\Sl_{\sigma'\in P(\{\beta_1\}\bigcup O\{\sigma\})}\mathcal{S}_{\{\beta_2,\beta_1,1,2,\alpha_1,...,\alpha_r,N\},\{1,\beta_1,2,\sigma,N\}}
\mathcal{S}_{\{\beta_1,1,2,\sigma,N\},\{1,2,\sigma',N\}}A_o(1,2,\sigma',N)\nn
&&+\Sl_{\tau\in P(O\{\beta_1,\beta_2\} \bigcup O\{\alpha\})}
\mathcal{S}_{\{\beta_2,\beta_1,1,2,\alpha_1,...,\alpha_r,N\},\{1,2,\tau,N\}}A_o(1,2,\tau,N)\Biggr],
\eea
where $s_{i_1i_2...i_j}=2\Sl_{1\leq m<n\leq j}k_{i_m}\cdot k_{i_n}$.
Considering different permutations of $\beta_1$ and $\beta_2$, we can express the above equation by
a sum over $\sigma\in P(O\{\alpha\}\bigcup\{\beta_1,\beta_2\})$.
If $\sigma^{-1}(\beta_1)>\sigma^{-1}(\beta_2)$,
 the second term of the above equation does not contribute to this permutation. The first term of
 \eqref{min-basis-e-2a} gives
\bea
\Sl_{\sigma\in P(O\{\alpha\}\in O\{\beta_1,\beta_2\})}\left[-\frac{\mathcal{S}_{\{\beta_2,\beta_1,1,2,\alpha,N\},\{1,\beta_1,2,\sigma/\{\beta_1\},N\}}}{\sin(\pi\alpha's_{1\beta_1\beta_2})}\right]
\left[-\frac{\mathcal{S}_{\{\beta_1,1,2,\sigma/\{\beta_1\},N\},\{1,2,\sigma,N\}}}{\sin(\pi\alpha's_{1\beta_1}) }\right]A_o(1,2,\sigma,N),
\eea
 where $\sigma/\{\beta_1\}$ denotes the permutation of the legs except $\beta_1$ in $\sigma$.

If  $\sigma^{-1}(\beta_2)>\sigma^{-1}(\beta_1)$, both two terms of \eqref{min-basis-e-2a}
contribute to this kind of permutation
\bea
&&\Sl_{\sigma\in P(O\{\alpha\}\bigcup O\{\beta_2,\beta_1\})}\Biggl[\frac{\mathcal{S}_{\{\beta_2,\beta_1,1,2,\alpha,N\},\{1,\beta_1,2,\sigma/\{\beta_1\},N\}}}{\sin(\pi\alpha's_{1\beta_1\beta_2})}
\frac{\mathcal{S}_{\{\beta_1,1,2,\sigma/\{\beta_1\},N\},\{1,2,\sigma,N\}}}{\sin(\pi\alpha's_{1\beta_1}) }\nn
&&-\frac{\mathcal{S}_{\{\beta_2,\beta_1,1,2,\alpha,N\},\{1,2,\sigma,N\}}}{\sin(\pi\alpha's_{1\beta_1\beta_2})}\Biggr]A_o(1,2,\sigma,N).
\eea
Those in two different cases corresponding to different relative orderings of $\beta_1$ and $\beta_2$ in $\sigma$
can be written together
\bea
&&A_o(1,\beta_1,\beta_2,2,\alpha_1,...,\alpha_{N-5},N)\nn
&=&\Sl_{\sigma\in P(O\{\alpha\}\bigcup\{\beta_1,\beta_2\})}\Biggl[\frac{\mathcal{S}_{\{\beta_2,\beta_1,1,2,\alpha,N\},\{1,\beta_1,2,\sigma/\{\beta_1\},N\}}}{\sin(\pi\alpha's_{1\beta_1\beta_2})}
\frac{\mathcal{S}_{\{\beta_1,1,2,\sigma/\{\beta_1\},N\},\{1,2,\sigma,N\}}}{\sin(\pi\alpha's_{1\beta_1}) }\nn
&&-\frac{\mathcal{S}_{\{\beta_2,\beta_1,1,2,\alpha,N\},\{1,2,\sigma,N\}}}{\sin(\pi\alpha's_{1\beta_1\beta_2})}\theta(\sigma^{-1}(\beta_2)-\sigma^{-1}(\beta_1))\Biggr]A_o(1,2,\sigma,N).
\eea

{\bf Level-3}

The BCJ relations with three $\beta$s can be given as
\bea
&&A_o(1,\beta_1,\beta_2,\beta_3,2,\alpha_1,...,\alpha_{N-6},N)\nn
&=&-\frac{1}{\sin(\pi\alpha's_{1\beta_1\beta_2\beta_3})}\times\Biggl[\Sl_{\sigma\in P(O\{\beta_3\}\bigcup O\{\alpha\})}
\mathcal{S}_{\{\beta_3,\beta_2,\beta_1,1,2,\alpha,N\},\{1,\beta_1,\beta_2,2,\sigma,N\}}A_o(1,\beta_1,\beta_2,2,\sigma,N)\nn
&&+\Sl_{\sigma'\in P(O\{\beta_2,\beta_3\}\bigcup O\{\alpha\})}\mathcal{S}_{\{\beta_3,\beta_2,\beta_1,1,2,\alpha,N\},\{1,\beta_1,2,\sigma',N\}}A_o(1,\beta_1,2,\sigma',N)\nn
&&+\Sl_{\sigma''\in P(O\{\beta_1,\beta_2,\beta_3\}\bigcup O\{\alpha\})}\mathcal{S}_{\{\beta_3,\beta_2,\beta_1,1,2,\alpha,N\},\{1,2,\sigma'',N\}}A_o(1,2,\sigma'',N)\Biggr].
\eea
We notice that the second term only contributes to the permutations with the relative ordering $\sigma^{-1}(\beta_2)<\sigma^{-1}(\beta_3)$, thus we can multiply a theta function
$\theta(\sigma^{-1}(\beta_3)-\sigma^{-1}(\beta_2))$ to this term and then replace the permutations $\sigma'\in P(O\{\beta_2,\beta_3\}\bigcup O\{\alpha\})$ by $\sigma'\in P(\{\beta_2,\beta_3\}\bigcup O\{\alpha\})$.
Similarly, we can multiply $\theta(\sigma^{-1}(\beta_3)-\sigma^{-1}(\beta_2))\theta(\sigma^{-1}(\beta_2)-\sigma^{-1}(\beta_1))$ to the third term and then replace
$\sigma''\in P(O\{\beta_1,\beta_2,\beta_3\}\bigcup O\{\alpha\})$ by $\sigma''\in P(\{\beta_1,\beta_2,\beta_3\}\bigcup O\{\alpha\})$.
Substituting the minimal-basis expansions  at level-2 and level-1 into the first and the second term of this equation respectively,
we express every term in the above equation
 by summing over all permutations with the relative ordering of the $\alpha$s preserved
\bea\label{3pt_minimal}
&&A_o(1,\beta_1,\beta_2,\beta_3,2,\alpha_1,...,\alpha_{N-6},N)\nn
&=&\Sl_{\sigma \in P(O\{\alpha\}\bigcup\{\beta_1,\beta_2,\beta_3\})}\nn
&&\Biggl[(-1)^3\frac{\mathcal{S}_{\{\beta_3,\beta_2,\beta_1,1,2,\alpha,N\},\{1,\beta_1,\beta_2,2,\sigma/\{\beta_1,\beta_2\},N\}}}{\sin(i\pi\alpha's_{1\beta_1\beta_2\beta_3})}
\frac{\mathcal{S}_{\{\beta_2,\beta_1,1,2,\sigma/\{\beta_1,\beta_2\},N\},\{1,\beta_1,2,\sigma/\{\beta_1\},N\}}}{\sin(\pi\alpha's_{1\beta_1\beta_2})}\nn
&&\times\frac{\mathcal{S}_{\{\beta_1,1,2,\sigma/\{\beta_1\},N\},\{1,2,\sigma,N\}}}{\sin(\pi\alpha's_{1\beta_1}) }\nn
&&+\frac{\mathcal{S}_{\{\beta_3,\beta_2,\beta_1,1,2,\alpha,N\},\{1,\beta_1,\beta_2,2,\sigma/\{\beta_1,\beta_2\},N\}}}{\sin(\pi\alpha's_{1\beta_1\beta_2\beta_3})}
\frac{\mathcal{S}_{\{\beta_2,\beta_1,1,2,\sigma/\{\beta_1,\beta_2\},N\},\{1,2,\sigma,N\}}}{\sin(\pi\alpha's_{1\beta_1\beta_2})}\theta(\sigma^{-1}(\beta_2)-\sigma^{-1}(\beta_1))\nn
&&+\frac{\mathcal{S}_{\{\beta_3,\beta_2,\beta_1,1,2,\alpha,N\},\{1,\beta_1,2,\sigma/\{\beta_1\},N\}}}{\sin(\pi\alpha's_{1\beta_1\beta_2\beta_3})}
\theta(\sigma^{-1}(\beta_3)-\sigma^{-1}(\beta_2))\frac{\mathcal{S}_{\{\beta_1,1,2,\sigma/\{\beta_1\},N\},\{1,2,\sigma,N\}}}{\sin(\pi\alpha's_{1\beta_1})}\nn
&&-\frac{\mathcal{S}_{\{\beta_3,\beta_2,\beta_1,1,2,\alpha,N\},\{1,2,\sigma,N\}}}{\sin(\pi\alpha's_{1\beta_1\beta_2\beta_3})}
\theta(\sigma^{-1}(\beta_3)-\sigma^{-1}(\beta_2))\theta(\sigma^{-1}(\beta_2)-\sigma^{-1}(\beta_1))\Biggr]A_o(1,2,\sigma,N),\nn
\eea
where $\sigma/\{\beta_{i_1},...,\beta_{i_n}\}$ denotes the permutation of the legs in $\sigma$ except the legs
$\beta_{i_1}$,..., $\beta_{i_n}$. This is just the minimal-basis expansion with three $\beta$s.

\subsection{General formula}
It is easy to extend the discussions in the examples to general formula by induction. In general, we can divide
the ordered set $O\{\beta_1,...,\beta_s\}$ into $n(1\leq n\leq s)$ segments, i.e., $n$ ordered sets
$O\{\beta_1,...,\beta_{i_1}\}$, $O\{\beta_{i_1+1},...,\beta_{i_2}\}$,...,$O\{\beta_{i_{n-1}+1},...,\beta_s\}$.
We denote the last element of the $j$th ordered set as $\beta_{i_j}$. For any given division with $n$ segments,
we have $\beta_{i_n}=\beta_s$. We also define $i_0\equiv 0$ and the $0$th set is empty. For such a division with $n$ segments,
we get a contribution
\bea
\prod_{j=0}^{n-1}\left[-\frac{\mathcal{S}_{\{\beta_{i_{j+1}},...,\beta_1,1,2,\sigma/\{\beta_{i_{j+1}},...,\beta_1\},N\},\{1,\beta_1,...,\beta_{i_j},2,\sigma/\{\beta_1,...,\beta_{i_j}\},N\}}}
{\sin(\pi\alpha's_{1\beta_1,...,\beta_{i_j}})}
\Theta_{\sigma}(\beta_{i_j+1},...,\beta_{i_{j+1}})\right].
\eea
 $\Theta_{\sigma}(\beta_{i_j+1},...,\beta_{i_{j+1}})$ is defined as
\bea
\Theta_{\sigma}(\beta_{i_j+1},...,\beta_{i_{j+1}})=\Biggl\{
           \begin{array}{cc}
             \prod\limits_{i_{j}+1<k\leq i_{j+1}}(\sigma^{-1}(\beta_{k})-\sigma^{-1}(\beta_{k-1}))&(i_{j+1}\neq i_{j}+1) \\
             1&(i_{j+1}=i_{j}+1) \\
           \end{array}
\eea
where, the case with $i_{j+1}=i_{j}+1$ means there is only one element $\beta_{i_{j+1}}$.
Considering all the possible divisions for any given permutation
$\sigma$($\sigma\in P(O\{\alpha\}\bigcup\{\beta\})$), we can give the minimal-basis expansion as
\bea\label{minimal-basis}
&&A_o(1,\beta_1,...,\beta_s,2,\alpha_1,...,\alpha_{N-s-3},N)\nn
&=&\Sl_{\sigma\in P(O\{\alpha\}\bigcup\{\beta\})}\Sl_{\text{All divisions} O\{\beta\}\rightarrow O\{\beta_1,...,\beta_{i_1}\} O\{\beta_{i_1+1},...,\beta_{i_2}\},...,O\{\beta_{i_{n-1}+1},...,\beta_{i_n}\}}\nn
&&\prod_{j=0}^{n-1}\left[-\frac{\mathcal{S}_{\{\beta_{i_{j+1}},...,\beta_1,1,2,\sigma/\{\beta_{i_{j+1}},...,\beta_1\},N\},\{1,\beta_1,...,\beta_{i_j},2,\sigma/\{\beta_1,...,\beta_{i_j}\},N\}}}
{\sin(\pi\alpha's_{1\beta_1,...,\beta_{i_j}})}
\Theta_{\sigma}(\beta_{i_j+1},...,\beta_{i_{j+1}})\right]A_o(1,2,\sigma,N).\nn
\eea
This can be seen from the examples explicitly, e.g., for the case with three $\beta$s, there are four possible divisions of $O\{\beta_1,\beta_2,\beta_3\}$. The first division is
$O\{\beta_1,\beta_2,\beta_3\}\rightarrow O\{\beta_1\},O\{\beta_2\},O\{\beta_3\}$. The second division is
$O\{\beta_1,\beta_2,\beta_3\}\rightarrow O\{\beta_1,\beta_2\}$, $O\{\beta_3\}$. The third division is
$O\{\beta_1,\beta_2,\beta_3\}\rightarrow O\{\beta_1\},\{\beta_2,\beta_3\}$ and the fourth division is
$O\{\beta_1,\beta_2,\beta_3\}\rightarrow O\{\beta_1,\beta_2,\beta_3\}$. The contributions to the minimal-basis
expansion of this four divisions just corresponds to the four terms in \eqref{3pt_minimal}.

To prove this minimal-basis expansion, we should consider the BCJ relation with $s$ $\beta$s. Using the BCJ relation with $s$ $\beta$s, we obtain
\bea
&&A_o(1,\beta_1,...,\beta_s,2,\alpha_1,...,\alpha_{N-s-3},N)\nn
&=&-\frac{1}{\sin(\pi\alpha's_{1\beta_1...\beta_s})}
\Biggl[\Sl_{\sigma\in P(O\{\beta_{1},...,\beta_s\}\bigcup O\{\alpha\})}\mathcal{S}_{\{\beta^T,1,2,\alpha,N\},\{1,2,\sigma,N\}}A_o(1,2,\sigma,N)\nn
&+&\Sl_{l=1}^{s-1}\Sl_{\sigma\in P(O\{\beta_{l+1},...,\beta_s\}\bigcup O\{\alpha\})}\mathcal{S}_{\{\beta^T,1,2,\alpha,N\},\{1,\beta_1,...,\beta_l,2,\sigma,N\}}A_o(1,\beta_1,...,\beta_l,2,\sigma,N)\Biggr].
\eea
We can multiply  $\prod_{k=2}^{s}\theta(\sigma^{-1}(\beta_{k})-\sigma^{-1}(\beta_{k-1}))$ to the first term in the brackets
and sum over all $\sigma\in P(\{\beta_{1},...,\beta_s\}\bigcup O\{\alpha\})$ instead of $\sigma\in P(O\{\beta_{1},...,\beta_s\}\bigcup O\{\alpha\})$.
In a similar way, we can also multiply $\prod_{k=l+1}^{s}\theta(\sigma^{-1}(\beta_{k})-\sigma^{-1}(\beta_{k-1}))$ to the terms in the second line for any given
$l~(1\leq l< s-1)$ and use the sum over $\sigma\in P(\{\beta_{l+1},...,\beta_s\}\bigcup O\{\alpha\})$ instead of the sum over
$\sigma\in P(O\{\beta_{l+1},...,\beta_s\}\bigcup O\{\alpha\})$. The special case with $l=s-1$ only gets a trivial factor $1$ instead of the products of
theta functions.
After substituting the minimal-basis expansions with $\beta$s fewer than $s$, and noticing the definition of $\Theta$,  we get
\bea
&&A_o(1,\beta_1,...,\beta_s,2,\alpha_1,...,\alpha_{N-s-3},N)\nn
&=&
\Sl_{l=0}^{s-1}\Sl_{\sigma\in P(\{\beta_{l+1},...,\beta_s\}\bigcup O\{\alpha\})}\frac{-\mathcal{S}_{\{\beta^T,1,2,\alpha,N\},\{1,\beta_1,...,\beta_l,2,\sigma,N\}}}{\sin(\pi\alpha's_{1\beta_1...\beta_s})}
\Theta_{\sigma}(\beta_{l+1},...,\beta_{s})\nn
&\times&\Sl_{\sigma'\in P(O\{\sigma\}\bigcup\{\beta_1,...,\beta_l\})}\Sl_{\text{All divisions} O\{\beta_1,...,\beta_l\}\rightarrow O\{\beta_1,...,\beta_{i_1}\} O\{\beta_{i_1+1},...,\beta_{i_2}\},...,O\{\beta_{i_{n-1}+1},...,\beta_{l}\}}\nn
&&\prod_{j=0}^{n-1}\left[-\frac{\mathcal{S}_{\{\beta_{i_{j+1}},...,\beta_1,1,2,\sigma'/\{\beta_{i_{j+1}},...,\beta_1\},N\},\{1,\beta_1,...,\beta_{i_j},2,\sigma'/\{\beta_1,...,\beta_{i_j}\},N\}}}{\sin(\pi\alpha's_{1\beta_1,...,\beta_{i_j}})}
\Theta_{\sigma}(\beta_{i_j+1},...,\beta_{i_{j+1}})\right]A_o(1,2,\sigma',N).\nn
\eea
The two sums  $\Sl_{\sigma\in P(O\{\beta_{l+1},...,\beta_s\}\bigcup O\{\alpha\})}$ and $\Sl_{\sigma'\in P(O\{\sigma\}\bigcup\{\beta_1,...,\beta_l\})}$ means first merge
$\{\beta_{l+1},...,\beta_s\}$ with the ordered set $O\{\alpha\}$, then merge $\{\beta_1,...,\beta_l\}$ with $\{\beta_{l+1},...,\beta_s\}\bigcup O\{\alpha\}$. They can be written into only one sum
$\Sl_{\sigma\in P(\{\beta_{1},...,\beta_s\}\bigcup O\{\alpha\})}$. Since $\Sl_{\sigma\in P(O\{\beta_{1},...,\beta_s\}\bigcup O\{\alpha\})}$ is independent of $l$, it can commute with the sum $\Sl_{l=0}^{s-1}$. We have
\bea
&&A_o(1,\beta_1,...,\beta_s,2,\alpha_1,...,\alpha_{N-s-3},N)\nn
&=&
\Sl_{\sigma\in P(\{\beta_{1},...,\beta_s\}\bigcup O\{\alpha\})}\Sl_{l=0}^{s-1}\frac{-\mathcal{S}_{\{\beta^T,1,2,\sigma/\{\beta_1,...,\beta_s\},N\},
\{1,\beta_1,...,\beta_l,2,\sigma/\{\beta_1,...,\beta_l\},N\}}}{\sin(\pi\alpha's_{1\beta_1...\beta_s})}
\Theta_{\sigma}(\beta_{l+1},...,\beta_{s})\nn
&\times&\Sl_{\text{All divisions} O\{\beta_1,...,\beta_l\}\rightarrow O\{\beta_1,...,\beta_{i_1}\} O\{\beta_{i_1+1},...,\beta_{i_2}\},...,O\{\beta_{i_{n-1}+1},...,\beta_{l}\}}\nn
&&\prod_{j=0}^{n-1}\left[-\frac{\mathcal{S}_{\{\beta_{i_{j+1}},...,\beta_1,1,2,\sigma/\{\beta_{i_{j+1}},...,\beta_1\},N\},
\{1,\beta_1,...,\beta_{i_j},2,\sigma/\{\beta_1,...,\beta_{i_j}\},N\}}}{\sin(\pi\alpha's_{1\beta_1,...,\beta_{i_j}})}
\Theta_{\sigma}(\beta_{i_j+1},...,\beta_{i_{j+1}})\right]A_o(1,2,\sigma,N).\nn
\eea
Here we sum over $\sigma\in P(O\{\beta_{l+1},...,\beta_s\}\bigcup O\{\alpha\})$, all the sets in the momentum kernel are expressed by $\sigma$ modulo  a subset of $\{\beta_1,...,\beta_s\}$.
For example we express $\alpha$ by $\sigma/\{\beta_1,...,\beta_s\}$.
We can consider the ordered set $O\{\beta_{l+1},...,\beta_s\}$ as the last set in a division of $O\{\beta_1,...,\beta_s\}$.
Different $l$ corresponds to the divisions with different number of elements in the last set. Together with the divisions $O\{\beta_1,...,\beta_l\}\rightarrow O\{\beta_1,...,\beta_{i_1}\} O\{\beta_{i_1+1},...,\beta_{i_2}\},...,O\{\beta_{i_{n-1}+1},...,\beta_{l}\}$,
this gives all the possible divisions of $O\{\beta_{1},...,\beta_s\}$.  At last, we get the minimal-basis expansion \eqref{minimal-basis}.

\section{Conclusion}\label{section_6}
In this paper, we show that there are two primary relations among all the relations for open string tree amplitudes.
One of the primary relations can be chosen as the cyclic symmetry, the other one can be chosen as either the fundamental
KK relation or the fundamental BCJ relation. In field theory, the primary relations can only be chosen as the fundamental
BCJ relation and the cyclic symmetry. We establish a kind of general monodromy relation which can also be generated by the primary
relations. The general formula of the explicit minimal-basis expansions for open string tree amplitudes is given in this paper.

\subsection*{Acknowledgements}
This work is supported in part by the
NSF of China Grant No. 11105118, No. 10775116, No. 11075138, and 973-Program Grant
No. 2005CB724508.



\begin{thebibliography}{999}

\bibitem{Kleiss:1988ne}
  R.~Kleiss and H.~Kuijf,
  ``MULTI - GLUON CROSS-SECTIONS AND FIVE JET PRODUCTION AT HADRON COLLIDERS,''
  Nucl.\ Phys.\  B {\bf 312} (1989) 616.


\bibitem{Bern:2008qj}
  Z.~Bern, J.~J.~M.~Carrasco and H.~Johansson,
  ``New Relations for Gauge-Theory Amplitudes,''
  Phys.\ Rev.\  D {\bf 78} (2008) 085011
  [arXiv:0805.3993 [hep-ph]].

\bibitem{Kawai:1985xq}
  H.~Kawai, D.~C.~Lewellen and S.~H.~H.~Tye,
  ``A Relation Between Tree Amplitudes of Closed and Open Strings,''
  Nucl.\ Phys.\  B {\bf 269} (1986) 1.



\bibitem{BjerrumBohr:2009rd}
  N.~E.~J.~Bjerrum-Bohr, P.~H.~Damgaard and P.~Vanhove,
  ``Minimal Basis for Gauge Theory Amplitudes,''
  Phys.\ Rev.\ Lett.\  {\bf 103} (2009) 161602
  [arXiv:0907.1425 [hep-th]].
\bibitem{Stieberger:2009hq}
  S.~Stieberger,
  ``Open \& Closed vs. Pure Open String Disk Amplitudes,''
  arXiv:0907.2211 [hep-th].
\bibitem{BjerrumBohr:2010zs}
  N.~E.~J.~Bjerrum-Bohr, P.~H.~Damgaard, T.~Sondergaard and P.~Vanhove,
  ``Monodromy and Jacobi-like Relations for Color-Ordered Amplitudes,''
  JHEP {\bf 1006} (2010) 003
  [arXiv:1003.2403 [hep-th]].



\bibitem{BjerrumBohr:2010hn}
  N.~E.~J.~Bjerrum-Bohr, P.~H.~Damgaard, T.~Sondergaard and P.~Vanhove,
  ``The Momentum Kernel of Gauge and Gravity Theories,''
  JHEP {\bf 1101} (2011) 001
  [arXiv:1010.3933 [hep-th]].
\bibitem{DelDuca:1999rs}
  V.~Del Duca, L.~J.~Dixon and F.~Maltoni,
  ``New color decompositions for gauge amplitudes at tree and loop level,''
  Nucl.\ Phys.\  B {\bf 571} (2000) 51
  [arXiv:hep-ph/9910563].




\bibitem{Feng:2010my}
  B.~Feng, R.~Huang and Y.~Jia,
  ``Gauge Amplitude Identities by On-shell Recursion Relation in S-matrix
  Program,''
  Phys.\ Lett.\  B {\bf 695} (2011) 350
  [arXiv:1004.3417 [hep-th]].

\bibitem{Jia:2010nz}
  Y.~Jia, R.~Huang and C.~Y.~Liu,
  ``$U(1)$-decoupling, KK and BCJ relations in $\mathcal{N}=4$ SYM,''
  Phys.\ Rev.\  D {\bf 82} (2010) 065001
  [arXiv:1005.1821 [hep-th]].
\bibitem{Britto:2004ap}
  R.~Britto, F.~Cachazo and B.~Feng,
  ``New recursion relations for tree amplitudes of gluons,''
  Nucl.\ Phys.\  B {\bf 715} (2005) 499
  [arXiv:hep-th/0412308].

\bibitem{Britto:2005fq}
  R.~Britto, F.~Cachazo, B.~Feng and E.~Witten,
  ``Direct proof of tree-level recursion relation in Yang-Mills theory,''
  Phys.\ Rev.\ Lett.\  {\bf 94} (2005) 181602
  [arXiv:hep-th/0501052].
\bibitem{Tye:2010kg}
  H.~Tye and Y.~Zhang,
  ``Comment on the Identities of the Gluon Tree Amplitudes,''
  arXiv:1007.0597 [hep-th].



\bibitem{Chen:2011jx}
  Y.~X.~Chen, Y.~J.~Du and B.~Feng,
  ``A Proof of the Explicit Minimal-basis Expansion of Tree Amplitudes in Gauge
  Field Theory,''
  JHEP {\bf 1102} (2011) 112
  [arXiv:1101.0009 [hep-th]].
\bibitem{BjerrumBohr:2010ta}
  N.~E.~J.~Bjerrum-Bohr, P.~H.~Damgaard, B.~Feng and T.~Sondergaard,
  ``Gravity and Yang-Mills Amplitude Relations,''
  Phys.\ Rev.\  D {\bf 82} (2010) 107702
  [arXiv:1005.4367 [hep-th]].
\bibitem{BjerrumBohr:2010zb}
  N.~E.~J.~Bjerrum-Bohr, P.~H.~Damgaard, B.~Feng and T.~Sondergaard,
  ``New Identities among Gauge Theory Amplitudes,''
  Phys.\ Lett.\  B {\bf 691} (2010) 268
  [arXiv:1006.3214 [hep-th]].
\bibitem{BjerrumBohr:2010yc}
  N.~E.~J.~Bjerrum-Bohr, P.~H.~Damgaard, B.~Feng and T.~Sondergaard,
  ``Proof of Gravity and Yang-Mills Amplitude Relations,''
  JHEP {\bf 1009} (2010) 067
  [arXiv:1007.3111 [hep-th]].
\bibitem{Feng:2010br}
  B.~Feng and S.~He,
  ``KLT and New Relations for N=8 SUGRA and N=4 SYM,''
  JHEP {\bf 1009} (2010) 043
  [arXiv:1007.0055 [hep-th]].



\bibitem{Bern:2010ue}
  Z.~Bern, J.~J.~M.~Carrasco and H.~Johansson,
  ``Perturbative Quantum Gravity as a Double Copy of Gauge Theory,''
  Phys.\ Rev.\ Lett.\  {\bf 105} (2010) 061602
  [arXiv:1004.0476 [hep-th]].
\bibitem{Bern:2010yg}
  Z.~Bern, T.~Dennen, Y.~t.~Huang and M.~Kiermaier,
  ``Gravity as the Square of Gauge Theory,''
  Phys.\ Rev.\  D {\bf 82} (2010) 065003
  [arXiv:1004.0693 [hep-th]].


\bibitem{Mafra:2011kj}
  C.~R.~Mafra, O.~Schlotterer and S.~Stieberger,
  ``Explicit BCJ Numerators from Pure Spinors,''
  arXiv:1104.5224 [hep-th].


\bibitem{Monteiro:2011pc}
  R.~Monteiro and D.~O'Connell,
  ``The Kinematic Algebra From the Self-Dual Sector,''
  arXiv:1105.2565 [hep-th].


\bibitem{Tye:2010dd}
  S.~H.~Henry Tye and Y.~Zhang,
  ``Dual Identities inside the Gluon and the Graviton Scattering Amplitudes,''
  JHEP {\bf 1006} (2010) 071
  [Erratum-ibid.\  {\bf 1104} (2011) 114]
  [arXiv:1003.1732 [hep-th]].

\bibitem{Mafra:2009bz}
  C.~R.~Mafra,
  ``Simplifying the Tree-level Superstring Massless Five-point Amplitude,''
  JHEP {\bf 1001} (2010) 007
  [arXiv:0909.5206 [hep-th]].

\bibitem{Mafra:2010gj}
  C.~R.~Mafra, O.~Schlotterer, S.~Stieberger and D.~Tsimpis,
  ``Six Open String Disk Amplitude in Pure Spinor Superspace,''
  Nucl.\ Phys.\  B {\bf 846} (2011) 359
  [arXiv:1011.0994 [hep-th]].



\bibitem{Mafra:2011nv}
  C.~R.~Mafra, O.~Schlotterer and S.~Stieberger,
  ``Complete N-Point Superstring Disk Amplitude I. Pure Spinor Computation,''
  arXiv:1106.2645 [hep-th].




\bibitem{Mafra:2011nw}
  C.~R.~Mafra, O.~Schlotterer and S.~Stieberger,
  ``Complete N-Point Superstring Disk Amplitude II. Amplitude and
  Hypergeometric Function Structure,''
  arXiv:1106.2646 [hep-th].




\bibitem{Chen:2010ct}
  Y.~X.~Chen, Y.~J.~Du and B.~Feng,
  ``On tree amplitudes with gluons coupled to gravitons,''
  JHEP {\bf 1101} (2011) 081
  [arXiv:1011.1953 [hep-th]].

\bibitem{Sondergaard:2009za}
  T.~Sondergaard,
  ``New Relations for Gauge-Theory Amplitudes with Matter,''
  Nucl.\ Phys.\  B {\bf 821} (2009) 417
  [arXiv:0903.5453 [hep-th]].
\bibitem{Du:2011js}
  Y.~J.~Du, B.~Feng and C.~H.~Fu,
  ``BCJ Relation of Color Scalar Theory and KLT Relation of Gauge Theory,''
  arXiv:1105.3503 [hep-th].

\bibitem{Feng:2009ei}
  B.~Feng, J.~Wang, Y.~Wang and Z.~Zhang,
  ``BCFW Recursion Relation with Nonzero Boundary Contribution,''
  JHEP {\bf 1001} (2010) 019
  [arXiv:0911.0301 [hep-th]].


\bibitem{Feng:2010ku}
  B.~Feng and C.~Y.~Liu,
  ``A Note on the boundary contribution with bad deformation in gauge theory,''
  JHEP {\bf 1007} (2010) 093
  [arXiv:1004.1282 [hep-th]].



\bibitem{Bern:1999bx}
  Z.~Bern, A.~De Freitas and H.~L.~Wong,
  ``On the coupling of gravitons to matter,''
  Phys.\ Rev.\ Lett.\  {\bf 84} (2000) 3531
  [arXiv:hep-th/9912033].


\bibitem{BjerrumBohr:2011xe}
  N.~E.~J.~Bjerrum-Bohr, P.~H.~Damgaard, H.~Johansson and T.~Sondergaard,
  ``Monodromy--like Relations for Finite Loop Amplitudes,''
  JHEP {\bf 1105} (2011) 039
  [arXiv:1103.6190 [hep-th]].
\bibitem{Feng:2011fja}
  B.~Feng, Y.~Jia and R.~Huang,
  ``Relations of loop partial amplitudes in gauge theory by Unitarity cut
  method,''
  arXiv:1105.0334 [hep-ph].
\bibitem{Bern:2011ia}
  Z.~Bern and T.~Dennen,
  ``A Color Dual Form for Gauge-Theory Amplitudes,''
  arXiv:1103.0312 [hep-th].

\bibitem{Del Duca:1999ha}
  V.~Del Duca, A.~Frizzo, F.~Maltoni,
  ``Factorization of tree QCD amplitudes in the high-energy limit and in the collinear limit,''
  Nucl.\ Phys.\  {\bf B568 } (2000)  211-262.
  [hep-ph/9909464].







\end{thebibliography}
\end{document}